\documentclass[paper,notoc]{JHEP3}
\pdfoutput=1
\usepackage{bbm}
\usepackage{amsmath}
\usepackage{graphicx,subfig}
\usepackage{amssymb}

\title{Modified Gravity: the CMB, Weak Lensing and General Parameterisations}
\author{Shaun A. Thomas$^{1}$, Stephen A. Appleby$^{2,3}$, Jochen Weller$^{2,3,4}$ \\
$^{1}$Department of Physics and Astronomy, University College London, Gower Street, London, WC1E 6BT, UK \\
$^{2}$Excellence Cluster Universe, Boltzmannstr. 2, 85748 Garching, Germany \\
$^{3}$University Observatory, Ludwig-Maximillians University Munich,  \\ Scheinerstr. 1, 81679 Munich, Germany \\ 
$^{4}$Max-Planck-Institut f\"{u}r extraterrestrische Physik, \\ Giessenbachstrasse, 85748 Garching, Germany \\
Email: \email{shaun.thomas@ucl.ac.uk}, \email{stephen.appleby@ph.tum.de}, \email{jochen.weller@usm.lmu.de}}

\preprint{arXiv:}

\date{\today}

\abstract{We examine general physical parameterisations for viable gravitational models in the $f(R)$ framework. This is related to the mass of an additional scalar field, called the scalaron, that is introduced by the theories. Using a simple parameterisation for the scalaron mass $M(a)$ we show there is an exact correspondence between the model and popular parameterisations of the modified Poisson equation $\mu(a,k)$ and the ratio of the Newtonian potentials $\eta(a,k)$. However, by comparing the aforementioned model against other viable scalaron theories we highlight that the common form of $\mu(a,k)$ and $\eta(a,k)$ in the literature does not accurately represent $f(R)$ behaviour. We subsequently construct an improved description for the scalaron mass (and therefore $\mu(a,k)$ and $\eta(a,k)$) which captures their essential features and has benefits derived from a more physical origin. We study the scalaron's observational signatures and show the modification to the background Friedmann equation and CMB power spectrum to be small. We also investigate its effects in the linear and non linear matter power spectrum--where the signatures are evident--thus giving particular importance to weak lensing as a probe of these models. Using this new form, we demonstrate how the next generation Euclid survey will constrain these theories and its complementarity to current solar system tests. In the most optimistic case Euclid, together with a Planck prior, can constrain a fiducial scalaron mass $M_{0} = 9.4 \times 10^{-30}{\rm eV}$ at the $\sim 20 \%$ level. However, the decay rate of the scalaron mass, with fiducial value $\nu = 1.5$, can be constrained to $\sim 3\%$ uncertainty.}

\begin{document}

\section{Introduction}

Precision cosmological observations \cite{Spergel:2006hy,Komatsu:2008hk,Eisenstein:2005su,Riess:1998cb,Perlmutter:1998np} performed over the past decade provide overwhelming evidence that the Universe is currently undergoing a period of accelerated expansion. As a result the so-called `standard cosmological model' has emerged, which attributes the acceleration to a small but finite vacuum energy. Whilst this model is both simple and extremely successful, the fine tuning implicit in ensuring that the vacuum energy only dominates at low redshift has led to a proliferation of alternative `dark energy' models, which attempt to alleviate the fine tuning by introducing additional dynamical fields \cite{Copeland:2006wr}.

Another approach is to postulate that gravity deviates from its General Relativistic (GR) description on cosmological distance scales.  Perhaps the most common approach to modifying GR involves introducing additional fields, which mediate gravity in addition to the standard spin-2 graviton. In this work we consider one such class of models, $f(R)$ gravity, which is described by the following action

\begin{equation} \label{eq:1} S = \int \sqrt{-g} d^{4}x \left[ {R +f(R) \over 16\pi G} - {\cal L}_{\rm m} \right] , \end{equation}

\noindent where $f(R)$ is an unspecified function of the Ricci scalar, and ${\cal L}_{\rm m}$ is the Lagrange density of any non-gravitational fields. These models introduce an additional scalar field (hereafter referred to as the scalaron), which is gravitationally coupled to ordinary matter. This scalar field has the unique property that its mass depends on the `background' energy density via the so-called chameleon effect, a fact that proves important in evading solar system tests of gravity (see for example \cite{Chiba:2003ir,Hu:2007nk,Tsujikawa:2008uc,Lin:2010hk}). A significant body of literature has been devoted to models of the form ($\ref{eq:1}$), and we direct the reader to \cite{DeFelice:2010aj,Nojiri:2006ri,Capozziello:2007ec,Sotiriou:2008rp} for reviews on the subject.

In principle one can use an arbitrary function $f(R)$ in the action ($\ref{eq:1}$), however the vast majority of $f(R)$ models have been ruled out on either theoretical or observational grounds. Only a small subclass of functions are still considered `viable'.  In this work we consider the subset of models that can be written as expansions around General Relativity for all  $R > R_{\rm vac}$, where $R_{\rm vac} = 12H_{0}^{2}\Omega_{\Lambda}$ is the cosmological vacuum curvature at the present time. Examples of viable $f(R)$ functions are \cite{Hu:2007nk,st}

\begin{eqnarray} \label{eq:a1}& & f(R) =  -m^{2} {c_{1} (R/m^{2})^{2n} \over 1+ c_{2}(R/m^{2})^{2n}}  \\  \label{eq:a2} & & f(R)  =  \lambda R_{st} \left[ \left( 1 + {R^{2} \over R_{st}^{2}}\right)^{-d} - 1 \right]  \end{eqnarray}

\noindent both of which can be expanded as 

\begin{eqnarray} \label{eq:i1} & & f(R) =   -{R_{\rm vac} \over 2} + \lambda R_{\rm vac} \left({R_{\rm vac}  \over R} \right)^{2s} + {\cal O} \left( \lambda^{2} \left({R_{\rm vac}  \over R} \right)^{4s} \right)  \end{eqnarray}

\noindent for $R \gg R_{\rm vac}$, where $\lambda$ and $s$ are the modified gravity parameters nontrivially related to the model specific parameters $m^{2},c_{1},c_{2},R_{st}, n,d$ (see also \cite{ap,ts10,Cognola:2007zu,Nojiri:2006be,Fay:2007uy,Linder:2009jz,Bamba:2010ws,Nojiri:2008nt,Park:2010da,Elizalde:2010ts,Appleby:2009uf}). The fact that these functions reduce to expansions around a constant in regions of high curvature allow them to evade solar system tests and also reproduce the standard background cosmology with radiation and matter dominated epochs \cite{Amendola:2006kh,Amendola:2006we}.\footnote{The functions ($\ref{eq:a1},\ref{eq:a2}$)  in fact still possess potential fine tuning issues in the very early Universe, see for example \cite{st}. However, we only consider redshifts $z \lesssim 1000$ and assume that these issues are ameliorated by modifying the $f(R)$ functional form at high energies.}

Observational constraints on $f(R)$ models place lower bounds on the mass of the scalar field, at the curvature scale for which the observation is performed. In this work we will be concerned with cosmological constraints that can be placed on $f(R)$ models and associated modified gravity parameterisations. One can broadly classify cosmological probes in two categories: distance based probes (such as SNIa), which utilize the background evolution of the Hubble parameter, and structure based probes (such as weak lensing and galaxy clustering), which require knowledge of the evolution and scale dependence of density perturbations.
\\
\\
\noindent The paper proceeds as follows: 
\\
\\
In section \ref{sec:1} we review the background evolution and perturbation equations for $f(R)$ models under the so called `quasi-static' approximation and explore a simple parameterisation to encapsulate them. In section \ref{sec:paramaterisations} we show that the scalaron parameterisation that we initially describe is equivalent to common parameterisations of the Poisson equation $\mu(a,k)$ and anisotropic stress $\eta(a,k)$. Then, by comparing between scalaron models we highlight these forms to be less flexible than at first thought. We therefore suggest an improved parameterisation, which accurately represents viable $f(R)$ models in the literature. We discuss potential modified gravity signals in the expansion history in \ref{sec:lum}, and in section \ref{sec:3} we consider modified gravity effects in the CMB angular power spectrum and the linear and non linear matter power spectrum.  In section \ref{sec:4} and \ref{sec:5} we examine how future CMB and weak lensing surveys (which probe the matter power spectrum) will constrain such models and argue for their complementarity to solar system constraints. We discuss our analysis and conclude in section \ref{sec:6}. 

\section{The Field Equations and Scalaron Mass Parameterisation} \label{sec:1}

We begin by briefly reviewing the modified gravity framework and parameterisations that we will be using in the paper (see \cite{Appleby:2010dx} for further details). The  $f(R)$ field equations are a system of fourth order differential equations for the metric, however for models which reduce to expansions around General Relativity one can use the quasi-static approximation  \cite{Boisseau:2000pr,Zhang:2005vt} to simplify them considerably.\footnote{The quasi-static approximation reduces the order of the field equations, and as such we will be discarding certain solutions. However it can be shown that the additional propagating degree of freedom in a cosmological background is an oscillatory mode, which will decay in the early Universe and can be set to zero in what follows (see \cite{st,Appleby:2008tv} for a discussion of this point).} The quasi-static approximation is applicable whenever the scalar field mass $M$ is large relative to the background expansion $H_{\rm background}$: $M^{2} \gg H^{2}_{\rm background}$. All current constraints on $f(R)$ models \cite{Song:2007da,Lombriser:2010mp,Schmidt:2009am,Smith:2009fn} indicate that our use of this approximation scheme is justified, however we are also careful to check retrospectively that any parameter constraints we project for these models respect the assumption $M^{2} \gg H^{2}_{\rm background}$. We do not discount the possibility that more exotic models might be constructed which violate the quasi-static approximation, however we do not consider them here.

In a cosmological context, the quasi-static approximation posits that the mass of the scalar field $M(a)$ (which will be time dependent in a FRW spacetime) satisfies $M^{2}(a) \gg H^{2}(a)$ for all $a$. For $M^{2}(a) \gg H^{2}(a)$ the modified Friedmann equation can be written as

\begin{equation}\label{eq:e3} {H^{2} \over H_{0}^{2}} = {\Omega_{m0} \over a^{3}} + {\Omega_{\rm r0} \over a^{4}} + \Omega_{\Lambda 0} + {R_{\rm GR} \over 6H_{0}^{2}}\bar{f}_{\rm R} - {\bar{f} \over 6H_{0}^{2}} - a{H_{\rm GR}^{2} \over H_{0}^{2}} R'_{\rm GR}\bar{f}_{\rm RR} - {H_{\rm GR}^{2} \over H_{0}^{2}}\bar{f}_{\rm R}, \end{equation}

\noindent where $H_{\rm GR}$ is the standard $\Lambda$CDM Hubble parameter

\begin{equation} {H_{\rm GR}^{2} \over H_{0}^{2}} = {\Omega_{m0} \over a^{3}} + \Omega_{\Lambda 0} + {\Omega_{\rm r0} \over a^{4}} ,\end{equation}

\noindent and in what follows, all $f,f_{\rm R}, f_{\rm RR}$ are functions of $R_{\rm GR} = 6\dot{H}_{\rm GR} + 12 H_{\rm GR}^{2}$. $R$ subscripts denote differentiation with respect to $R$; $f_{R} = df/dR$, $f_{RR} = d^{2}f/dR^{2}$ and we have defined $\bar{f} \equiv f + R_{\rm vac}/2$. Note also that we are assuming zero curvature, $\Omega_{k0} = 0$. In the quasi-static approximation one can relate the {\it scalaron mass} to the functional form of $\bar{f}$ via

\begin{equation}\label{eq:mass} M^{2} = {1 \over 3 \bar{f}_{RR}}. \end{equation}

\noindent It follows that the last four terms on the right hand side of ($\ref{eq:e3}$), which constitute the effect of the scalaron on the background evolution, are suppressed by order ${\cal O} (H_{\rm GR}^{2}/M^{2}) \ll 1$ relative to the standard GR contributions. In this way one expects $w \approx -1$.

The perturbation equations \cite{Bean:2006up} are also simplified when using the quasi-static approximation. We consider the scalar perturbations of the metric in the Newtonian gauge

\begin{equation} ds^{2} = -(1+2\psi)dt^{2} + a^{2}(t)(1-2\phi)\gamma_{ij}dx^{i}dx^{j} . \end{equation}

\noindent Considering the evolution of subhorizon modes $k \gg a H$, we can neglect all terms of order $H^{2}/M^{2}$ and take $k\phi /a, k\psi/a \gg \dot{\phi}, \dot{\psi}$, in which case the linearized equations reduce to 

\begin{eqnarray} \label{eq:16} & &  \psi = \left(1 + {2\bar{K}^{2} \over 3 + 2\bar{K}^{2}}\right) \phi , \\ \label{eq:17} & & k^{2}\phi = -4\pi G \left({3 + 2\bar{K}^{2} \over 3 + 3\bar{K}^{2}}\right)a^{2}\rho_{\rm m} \delta_{\rm m} , \\ \label{eq:18}  & & \label{eq:p1} \ddot{\delta}_{\rm m} + 2H \dot{\delta}_{\rm m} - 4\pi G \left({3 + 4\bar{K}^{2} \over 3 + 3\bar{K}^{2}}\right)\rho_{\rm m}\delta_{\rm m} = 0 , \end{eqnarray}

\noindent where $\bar{K} \equiv k/aM(a)$. Superhorizon modes will evolve according to General Relativity, as they satisfy $k_{\rm superh} \ll a M(a)$ at all times. From equations ($\ref{eq:16}-\ref{eq:18}$) it is clear that the scalaron introduces a redshift and scale dependent modification to the anisotropic stress, Poisson equation and subsequently the growth of over-densities. It is in this way that the scalaron's mass can form a natural and physical parameterisation for deviations to gravity in the $f(R)$ framework. We discuss the relation to other common parameterisations in Section~\ref{sec:paramaterisations}. 

The mass of the scalar field can take numerous functional forms, however its general behaviour is typically very simple. For models that reduce to expansions around a cosmological constant, the mass can be written as a monotonic function that decays to the present. A simple functional form can be used to describe this behaviour;

\begin{equation} \label{eq:5} M^{2} = {M_{0}^{2} \over a^{2p}} , \end{equation}

\noindent  which contains two free parameters $M_{0}$ and $p$. These parameters represent the value of the mass at the present time and the rate at which $M(a)$ increases to the past, respectively. In principle one should introduce a cutoff in $M(a)$ such that it does not diverge as $a \to 0$, however this issue is of no concern for the redshifts considered. $\Lambda$CDM is recovered in the limit $M_{0} \to \infty$; where the scalaron mass becomes large and the field does not propagate. In this paper we write units for $M_{0}^{-1}$ in $[10^{28}{\rm h^{-1} \hspace{1mm} eV^{-1}}]$, for comparison $H_{0}^{-1}$ in these units is $H_{0}^{-1} = 7500 [10^{28}{\rm h^{-1} \hspace{1mm} eV^{-1}}]$. We take the boundary of the quasi-static limit to be $M^{-1}_{0} \lesssim 375  [10^{28}{\rm h^{-1} \hspace{1mm} eV^{-1}}]$, and $\Lambda$CDM is recovered if we set $M^{-1}_{0} = 0$. The only constraint that is imposed on the parameter $p$ is that the mass $M^{2}$ must grow to the past at a faster rate than the background curvature; $R/M^{2} \to 0$ for decreasing $a$. This forces one to choose $p > 2$, to ensure that $H^{2}/M^{2}$ decays during radiation domination. There is no upper bound and there is no reason to fix its value, however - the larger we take it the faster these models asymptote to General Relativity at early times. Typically one can consider reasonable fiducial values within the range $3 \le p \le 6$.  

\begin{figure}
  \begin{flushleft}
    \centering
    \begin{minipage}[c]{1.00\textwidth}
      \centering
      \includegraphics[width=7cm,height=7cm]{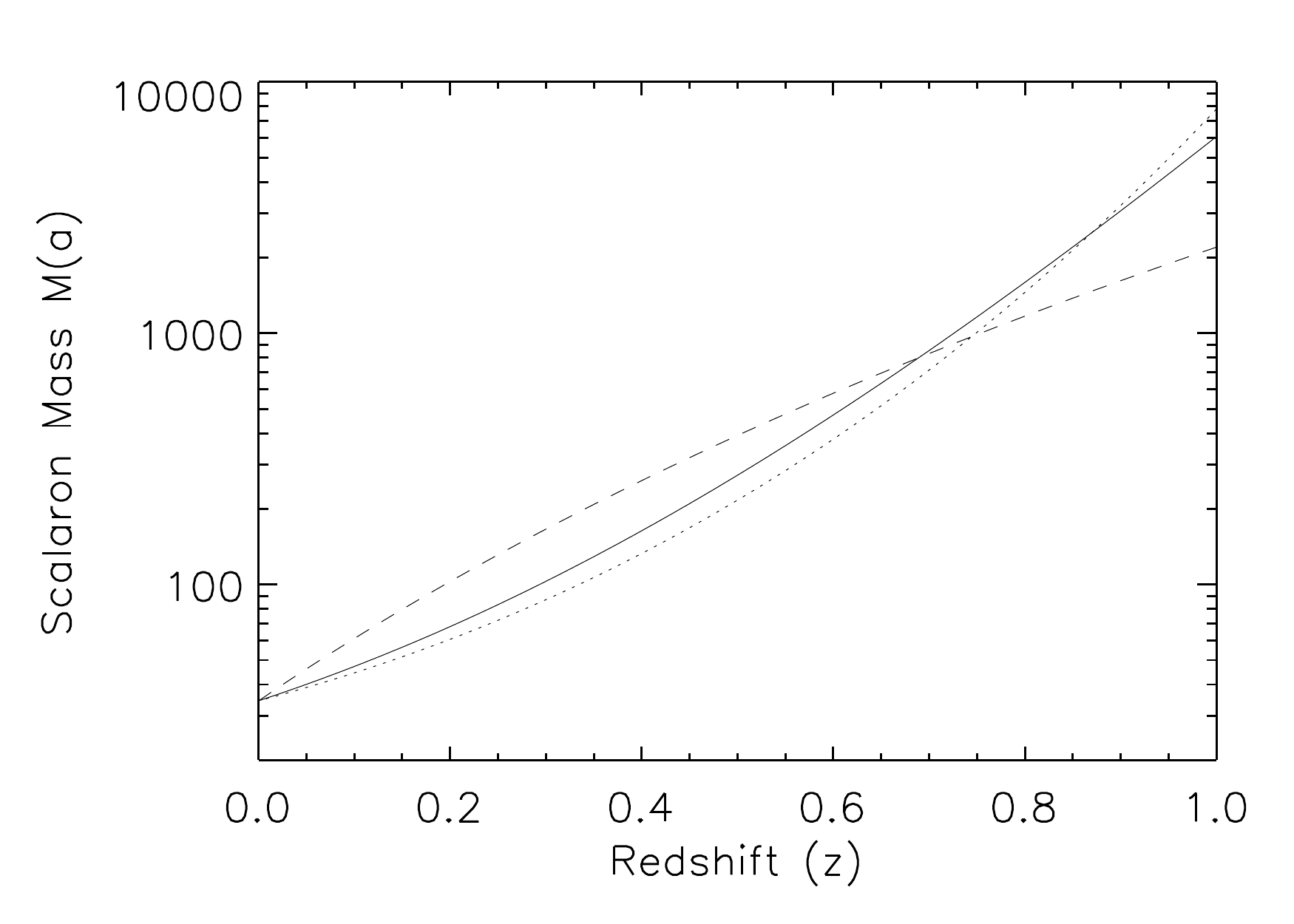}
    \end{minipage}
    \caption{\small{A comparison of different scalaron masses as a function of redshift: {\it simple} phenomenological (dashed; ($\ref{eq:5}$)), power law (solid; ($\ref{eq:i1}$)) and exponential (dotted; \cite{ap,Linder:2009jz,Bamba:2010ws}). They have been equated at $z=0$. The phenomenological model, which is exactly equivalent to parameterised deviations in the Poission equation and anisotropic stress, as shown in the next section, is not particularly adept at matching the other models. }}
    \label{fig:scalaronprofile}
  \end{flushleft}
\end{figure}

The expression ($\ref{eq:5}$) is chosen to approximately describe the evolution of the scalaron mass in a cosmological background. However, one should question how general this form is and how well it describes existing $f(R)$ functions in the literature. In Figure~\ref{fig:scalaronprofile} we compare the simple mass function ($\ref{eq:5}$) to a class of exponential models (dotted line) \cite{ap,Linder:2009jz,Bamba:2010ws}  and the power law model ($\ref{eq:i1}$; solid line) over $0<z<1$. It is clear that ($\ref{eq:5}$) does not accurately mimic the $f(R)$ models to which ($\ref{eq:i1}$) belongs at low redshift, and it is this regime that we are primarily interested. This is for two reasons: firstly, cosmological surveys are mainly sensitive to low and intermediate redshifts; and secondly, for redshifts beyond this the mass of the scalaron generically grows very rapidly, and hence any differences in $M(a)$ become irrelevant due to a highly suppressed modified gravity signal. Therefore, the exact form of $f(R)$ and hence $M(a)$ for higher redshifts is not particularly relevant; it is at this low redshift that we will observe an appreciable modified gravity signal.

The reason for the discrepancy is that the simple function fails to take into account the late time $\Lambda$ dominated epoch. During this period the Ricci scalar is approaching a constant future asymptote, and since $M = M[R(a)]$ it follows that the mass will also asymptote to a constant in this regime. To incorporate this behaviour we modify ($\ref{eq:5}$) according to

\begin{equation} \label{eq:107} M^{2} = M_{0}^{2} \left( { a^{-3} + 4 a_{*}^{-3} \over 1 + 4 a_{*}^{-3}}\right)^{2\nu} , \end{equation}

\noindent where we have defined $a_{*}$ as the scale factor at matter-$\Lambda$ equality: $a_{*} = (\Omega_{\rm m0}/\Omega_{\Lambda})^{1/3}$. It is important to note that we still have two modified gravity parameters: $M_{0}$ has the same interpretation as in ($\ref{eq:5}$) (the mass of the scalaron at the present, which must satisfy $M^{-1}_{0} \lesssim 375  [10^{28}{\rm h^{-1} \hspace{1mm} eV^{-1}}]$ in our approximation scheme) and $\nu$ is the rate of increase of $M(a)$ to the past, which we take to be $\nu > 3/2$. Note that $M(a)$ is now dependent on the background cosmology via $a_{*}$. The factors of four in ($\ref{eq:107}$) are included as the important scale is not $a_{*}$, but rather the scale at which $R$ transitions between matter and $\Lambda$ domination. Since  $R \propto \sum_{i}(\rho_{i} - 3P_{i})$, where $P_{\Lambda} = -\rho_{\Lambda}$ and $P_{m0} = 0$, the evolution of the Ricci scalar will be dominated by $\Lambda$ at an earlier time than $a_{*}$, hence the additional factor. This improved parameterization behaves like ($\ref{eq:5}$) in the matter dominated era (as expected), but has more appropriate behaviour at late times. It is this functional form that we will focus on for the majority of the paper. This parameterization is closely tied to the expansion ($\ref{eq:i1}$) and therefore can be considered as an effective parameterization and match to viable $f(R)$ models of the form ($\ref{eq:a1},\ref{eq:a2}$) considered in \cite{Hu:2007nk,st} and also the exponential models of \cite{ap,Linder:2009jz,Bamba:2010ws} in the redshift range $z \lesssim 1$ (shown in Figure~\ref{fig:scalaronprofile}).

\noindent

\section{Comparison to existing modified gravity parameterisations} \label{sec:paramaterisations}

Before we continue to analyse the new more suitable scalaron form ($\ref{eq:107}$) and its cosmological consequences we first examine the overly simplified parameterised mass function ($\ref{eq:5}$) in more detail. We find it is equivalent to an existing parameterisation in the literature which is widely used to represent modified gravity models, and highlights the limitations of that particular approach. We then discuss how our updated mass function $M(a)$ can be rewritten in terms of an improved metric potential parameterisation for those familiar with that formalism.

\subsection{Parameterisation Equivalence} \label{sec:correspondence}

In the quest to test deviations to gravity there have been many suggested methods and parameterisations. One popular procedure is to notice that in general the Poisson equation and metric potentials are altered \cite{Martinelli10,Bertschinger08,Zhao:2008bn,Daniel10} and so write

\begin{eqnarray} \label{eq:pa3} & &  k^{2} \psi = -4\pi G a^{2} \mu(a,k) \rho \delta_{\rm m} \\ \label{eq:4} & & {\phi \over \psi} = \eta(a,k). \end{eqnarray}

\noindent In particular \cite{Bertschinger08} first suggested that these two scale and redshift dependent functions take on forms to {\it specifically} reproduce the behaviour of $f(R)$ models:

\begin{eqnarray} \label{eq:pa5} & & \mu(a,k)  = {1 + \beta_{1} \lambda_{1}^{2} k^{2} a^{s} \over 1 + \lambda_{1}^{2} k^{2} a^{s}} \\ \label{eq:pa6} & & \eta(a,k) = {1 + \beta_{2} \lambda_{2}^{2} k^{2} a^{s} \over 1 + \lambda_{2}^{2} k^{2} a^{s}}. \end{eqnarray}

\noindent These have been further examined in \cite{Martinelli10,Zhao:2008bn,Giannantonio:2009gi} and have since become very popular. In this work we highlighted a simple functional form for the scalaron mass ($\ref{eq:5}$) to similarly represent $f(R)$ models. In the scalaron framework the physical scale dependence is implicitly specified by the redshift dependent mass: $\bar{K} \equiv k/ [a M(a)]$. We now exhibit the equivalence between these two approaches and obtain the relationship between the two sets of parameters $(\lambda_{1},s)$ and $(M_{0},p)$ in ($\ref{eq:5}$). To do so we write ($\ref{eq:16}$) as 

\begin{equation} \label{eq:pa7} \psi = \left(1 + {2\bar{K}^{2} \over 3 + 2\bar{K}^{2}}\right) \phi  = \left({ 3 + 4\bar{K}^{2} \over 3 + 2\bar{K}^{2}}\right) \phi  \end{equation}

\noindent so 
\begin{equation}\label{eq:pa8} {\phi \over \psi} = \left({ 3 + 2\bar{K}^{2} \over 3 + 4\bar{K}^{2}}\right). \end{equation}

\noindent The right hand side of ($\ref{eq:pa8}$) is clearly $\eta(a,k)$. Writing this out explicitly in terms of $M(a)$:

\begin{equation} \label{eq:pa9} { 3 + 2\bar{K}^{2} \over 3 + 4\bar{K}^{2}} = { 1 + {2 \over 3}\bar{K}^{2} \over 1 + {4 \over 3}\bar{K}^{2}} =  { 1 + {2 \over 3}\left({k^{2} \over M_{0}^{2}}\right) a^{2p-2} \over 1 + {4 \over 3} \left({k^{2} \over M_{0}^{2}}\right) a^{2p-2} } \end{equation}

\noindent one can now compare this to ($\ref{eq:pa6}$). Using the relations for scalar-tensor and $f(R)$ theories

\begin{equation} \beta_{1} = {4 \over 3} = {\lambda_{2}^{2} \over \lambda_{1}^{2}} = 2 - \beta_{2}{\lambda_{2}^{2} \over \lambda_{1}^{2}} \end{equation}

\noindent which are given in \cite{Martinelli10,Zhao:2008bn}, for example, we can immediately write down 

\begin{equation} \label{eq:restrictions}
\lambda_{2}^{2} = {4 \over 3} \lambda_{1}^{2}, \qquad \beta_{1} = {4 \over 3}, \qquad \beta_{2} = {1 \over 2}.
\end{equation}

\noindent Using these in ($\ref{eq:pa6}$) and equating with ($\ref{eq:pa9}$), we obtain the relationships
\begin{eqnarray} \label{eq:relation_1} & & s = 2p - 2 \\ \label{eq:relation_2} & & \lambda_{1} = {  1 \over M_{0}}. \end{eqnarray}

\noindent As a check, one can do the same process for equations ($\ref{eq:17}$) and ($\ref{eq:pa3}$), which again renders the same relations. Due to ($\ref{eq:restrictions}$) there are two free parameters in this $\mu(a,k)$ and $\eta(a,k)$ scheme: $s$ and $\lambda_{1}$; and so the two parameterisations are identical. Perhaps at first glance this is not so surprising -- the generalised forms ($\ref{eq:pa5}$) and ($\ref{eq:pa6}$) were invoked to match $f(R)$ behaviour. However, we previously showed that this simple model does not adequately describe general $f(R)$ behaviour and so neither do ($\ref{eq:pa5}$) and ($\ref{eq:pa6}$).

\subsection{Improved parameterisation} 

We now consider how the late time $\Lambda$ dominated epoch should be incorporated into the $\mu(a,k)$ and $\eta(a,k)$ parameterization. 

To this end we insert the mass function ($\ref{eq:107}$) into ($\ref{eq:16}$) and ($\ref{eq:17}$). In doing so we find that the $\Lambda$ dominated epoch can be accounted for by considering the following parameterized forms
\begin{eqnarray} \label{eq:o13} & & \mu(a,k)  = {1 + {4 \over 3} \lambda_{1}^{2} k^{2}  \beta(a) a^{s_{1}} \over 1 + \lambda_{1}^{2} k^{2} \beta(a) a^{s_{1}}} \\ \label{eq:o14} & & \eta(a,k) = {1 + {2 \over 3} \lambda_{1}^{2} k^{2} \beta(a)a^{s_{1}} \over 1 + {4 \over 3}\lambda_{1}^{2} k^{2} \beta(a) a^{s_{1}}}. \end{eqnarray}

\noindent where we have introduced the function $\beta(a)$, defined as

\begin{equation}\label{eq:o15} \beta(a) = \left[ a_{*}^{3} + 4a^{3}\right]^{-(s_{1}+2)/3}. \end{equation}

\noindent Again, there are still simply two free parameters: $(\lambda_{1},s_{1})$. The relationships between $(\lambda_{1},s_{1})$ and $(M_{0},\nu)$ in ($\ref{eq:107}$) are given by,
\begin{eqnarray} & & s_{1}=6\nu - 2 \\ & & \lambda_{1}^{2} = {(a_{*}^{3}+4)^{2\nu}
\over M_{0}^{2}}. \end{eqnarray}

\noindent This parameterisation is similar to the previous functional form at early times, however they differ at low redshifts due to $\beta(a)$. This is an important contribution, as it takes into account the fact that the mass of the scalar field typically asymptotes to a constant value for $a \sim 1$. Another advantage to using this form is its close link to physical $f(R)$ models of the form ($\ref{eq:a1}$)/($\ref{eq:a2}$). Current and novel approaches to modified gravity N-body simulations \cite{Oyaizu:2008tb,Zhao10} utilise an explicit form for $f(R)$ and therefore with a large suite of these simulations it will be possible to accurately reconstruct the nonlinear power spectrum for this more general gravitational parameterisation\footnote{Indeed these existing simulations have been for models resembling ($\ref{eq:a1}$) and ($\ref{eq:a2}$).}. 

In what follows we will focus on the parameterization of the scalaron mass given by the expression ($\ref{eq:107}$). Whilst these two improved approaches are equivalent, we regard the scalaron mass to be a quantity with a clear physical interpretation.

\section{Modified Gravity Signatures} \label{sec:2}

The system of equations ($\ref{eq:e3},\ref{eq:16}-\ref{eq:18}$) are sufficient to completely specify the background evolution and dynamics of the linear perturbations in viable $f(R)$ models, which are subject to the quasi-static approximation\footnote{Superhorizon modes are not accounted for in these equations, however they will simply evolve according to General Relativity.}. We now illustrate the typical modified gravity signals resulting from these equations using the phenomenological model ($\ref{eq:107}$). 

As described previously viable $f(R)$ models closely reproduce the $\Lambda$CDM expansion history, hence we expect only a very weak modified gravity signal in the Friedmann equation ($\ref{eq:e3}$) and a significantly stronger signal in the perturbation equations. This statement can be justified using a simple intuitive argument. The scalar field mass $M(a)$ introduces an additional scale into the field equations, in addition to the Hubble parameter $H(a)$ and the scale of the perturbations $k$. Perturbations deep inside the horizon satisfy $k^{2} \gg a^{2}H^{2}$, and we have chosen the mass such that $M^{2} \gg H^{2}$. However the relative size of $k$ and $M$ is undetermined. One can think of $M(a)$ as a `modified gravity horizon'; modes that satisfy $k^{2} < a^{2}M^{2}$ will evolve as they would under the standard General Relativistic field equations (with order $k^{2}/(a^{2}M^{2})$ corrections), whilst modes which satisfy $k^{2} > a^{2}M^{2}$ will exhibit order unity deviations from GR. It follows that a significant modified gravity signal will only be present above a particular (redshift dependent) scale $k_{\rm mg} = aM(a)$. This is clear from equations ($\ref{eq:e3}$,$\ref{eq:16}-\ref{eq:18}$); corrections to the standard Friedmann equation are of order ${\cal O}(H^{2}/M^{2})$), whilst the matter power spectrum exhibits large deviations from GR above a particular (redshift and model dependent) scale. We illustrate this in the next two subsections.

The above argument is supported in \cite{Martinelli:2009ek}, where constraints were imposed on viable $f(R)$ models using distance based cosmological probes. There it was found that only very weak constraints can be placed on modified gravity parameters, even when using several combined data sets.

\subsection{The Expansion History} \label{sec:lum}

\begin{figure*}
  \begin{flushleft}
    \centering
    \begin{minipage}[c]{1.00\textwidth}
      \centering
      \includegraphics[width=7cm,height=7cm]{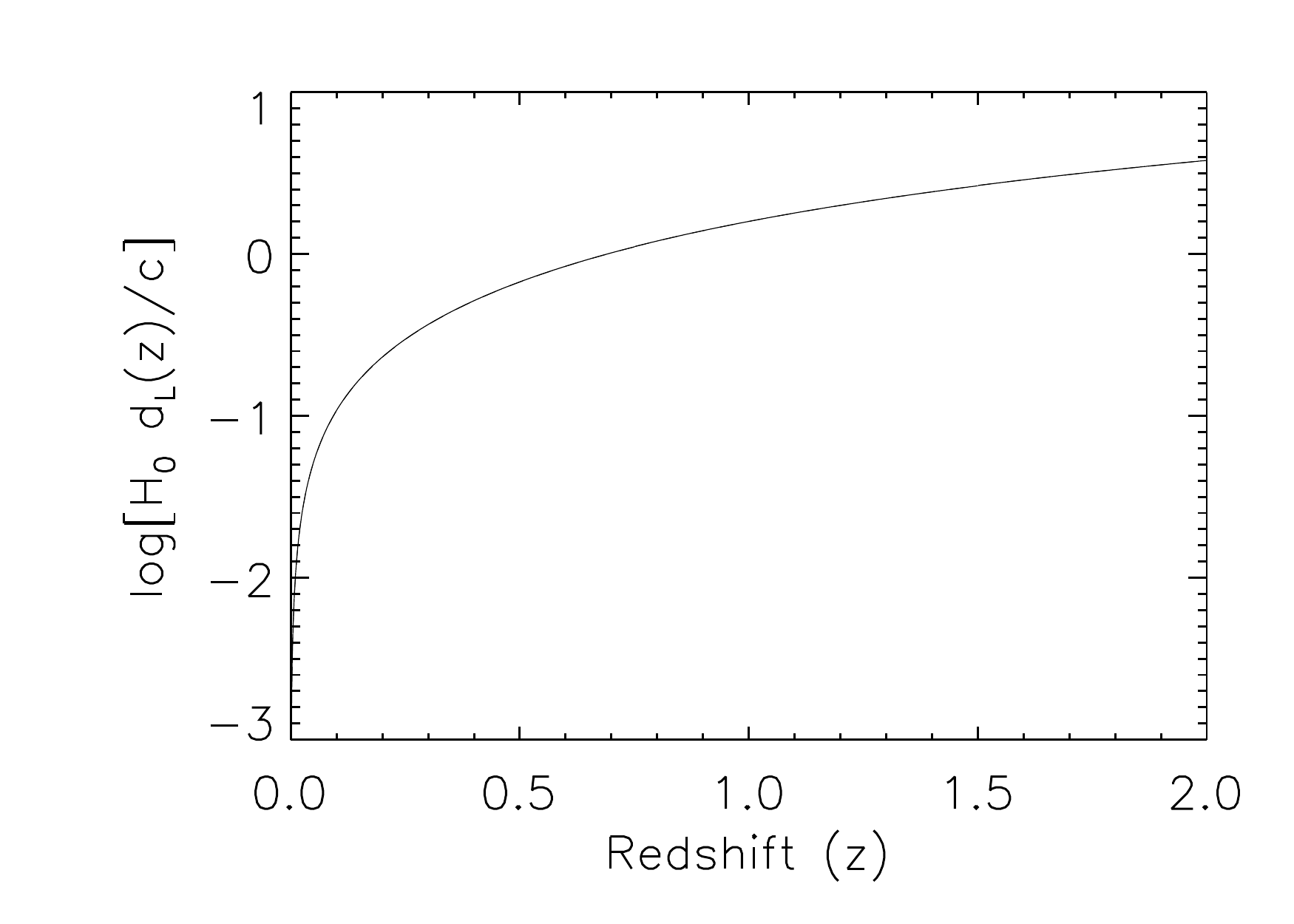}
      \includegraphics[width=7cm,height=7cm]{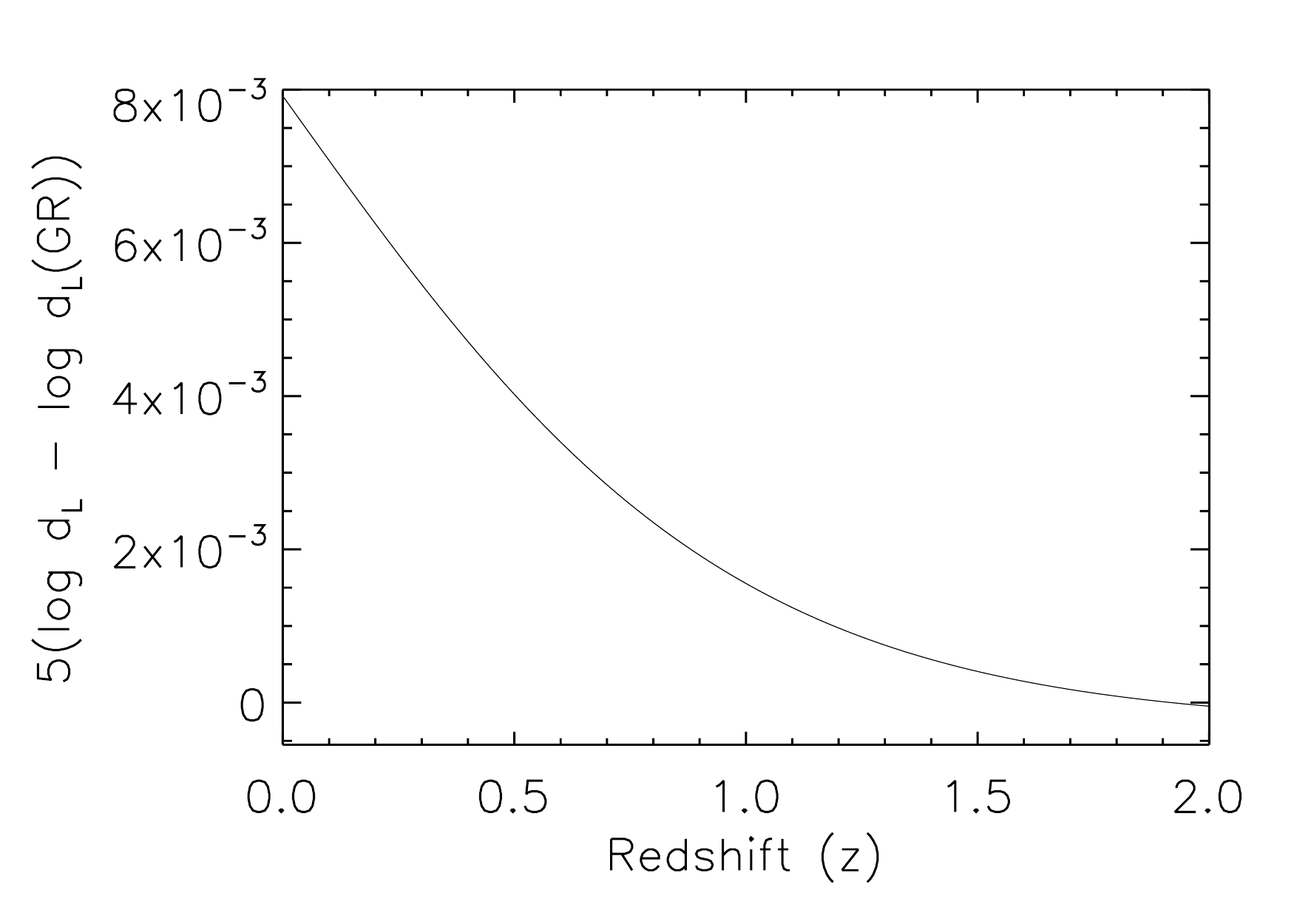}
    \end{minipage}
    \caption{\small{{\it Left Panel:} The luminosity distances for a $\Lambda$CDM background expansion history (solid line; $M^{-1}_{0}=0$, $\nu=1.5$) and the scalaron model (dashed line; $M^{-1}_{0}=375 [10^{28}{\rm h^{-1} \hspace{1mm} eV^{-1}}]$, $\nu=1.5$; indistinguishable). {\it Right Panel:} The fractional difference between the $\Lambda$CDM cosmology and parameterised deviation. It is clear from both panels that this framework closely mimics $\Lambda$CDM. }}
    \label{fig:1}
  \end{flushleft}
\end{figure*}
\noindent

We have argued that viable $f(R)$ models closely mimic the standard $\Lambda$CDM background expansion history. To confirm this, we take the general scalaron model (\ref{eq:107}) and compare the luminosity distance-redshift relations.

By substituting the mass $M(a)$ into the modified Friedmann equation ($\ref{eq:e3}$), we can construct the luminosity distance, defined in the usual way

\begin{equation} d_{\rm L} (z) = (1+z) \int_{0}^{z} {dz' \over H(z')} .\end{equation}

\noindent The results are exhibited in Figure~\ref{fig:1}, taking the fiducial parameters that we use throughout the paper: $\Omega_{m0} = 0.25$, $\Omega_{b0} = 0.05$, $h=0.7$, $\ln[A_{\rm s}] = \ln[2.34 \times 10^{-9}]$, $n_{\rm s} = 1$ and $\tau=0.9$. We contrast $M_{0}^{-1}=0.0$ and $\nu=1.5$ ($\Lambda$CDM) with $M^{-1}_{0}=375 [10^{28}{\rm h^{-1} \hspace{1mm} eV^{-1}}]$ and $\nu=1.5$ (the quasi-static limit), to show the maximum difference. We observe practically no signal for the modified gravity parameters chosen. This should come as no surprise; we have argued above that there will only be a relatively weak signal in the background cosmology for any model in the quasi-static approximation scheme. We can therefore effectively fix $w=-1$ and $w_{a}=0$ for the duration of the paper, defined as $w(a) = w + w_{a}(1-a)$. As we will see in the following section however, for the same parameter choices there is a significant modified gravity signal in the matter power spectrum.

\subsection{The Growth of Perturbations} \label{sec:3}

We now study the effect of the scalar field on the CMB angular power spectrum and matter power spectrum. In order to do so we modify {\sc camb} \cite{Lewis00} to incorporate the generalised modified gravity perturbation equations ($\ref{eq:16}-\ref{eq:p1}$), and use this to construct the modified source terms in the Boltzmann equations.

\begin{figure*}
  \begin{flushleft}
    \centering
    \begin{minipage}[c]{1.00\textwidth}
      \centering
      \includegraphics[width=7cm,height=7cm]{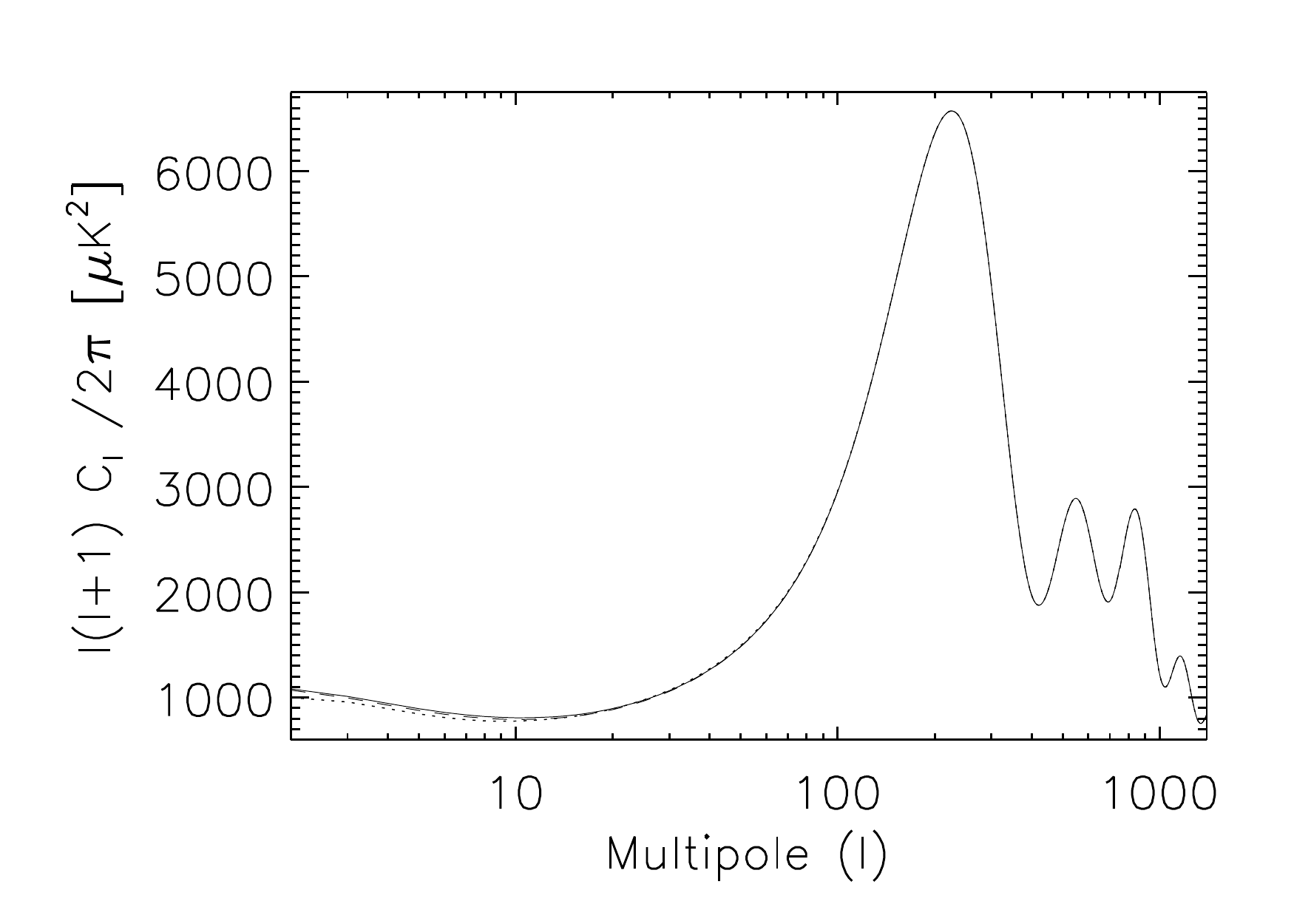}
      \includegraphics[width=7cm,height=7cm]{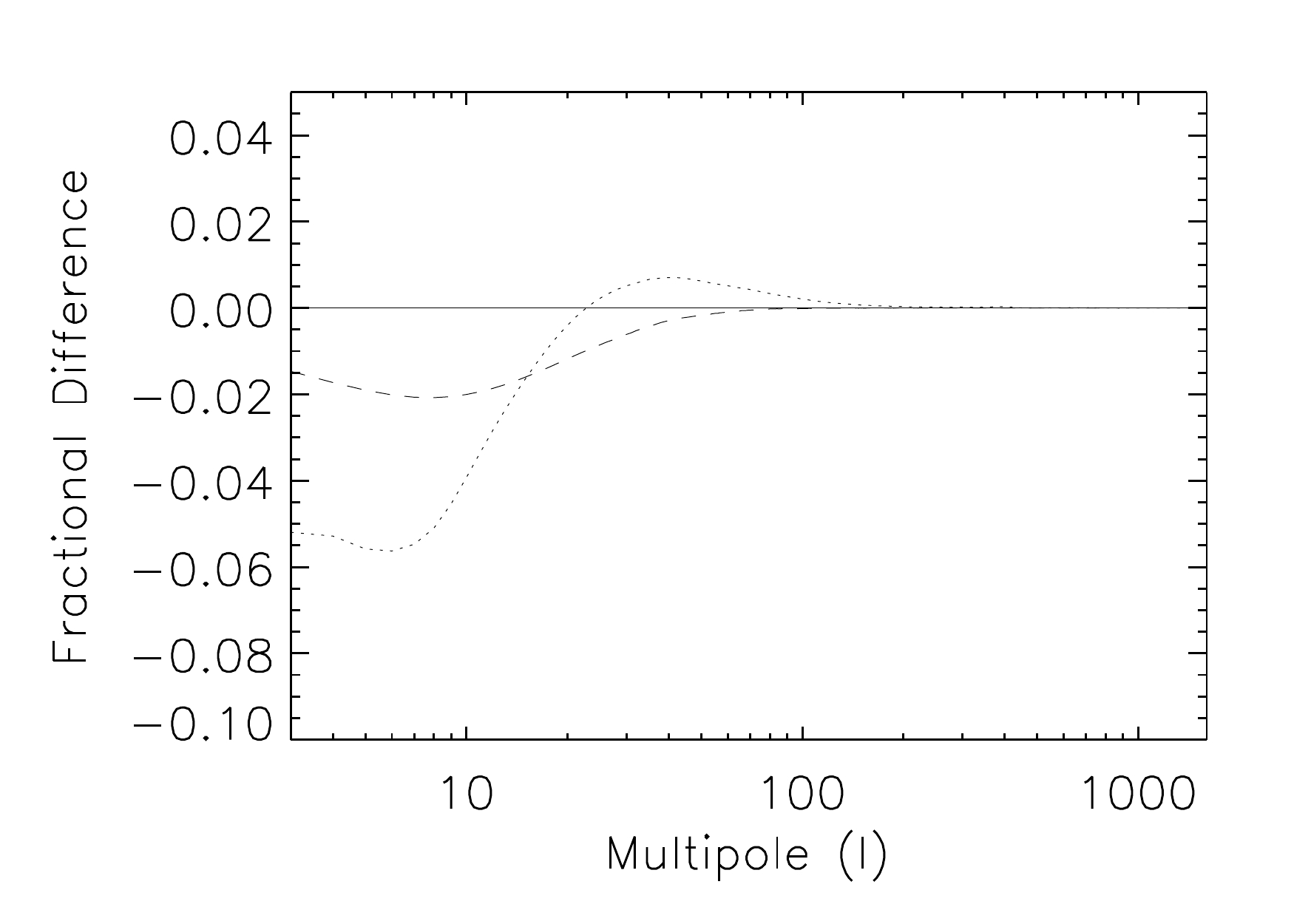}
    \end{minipage}
    \caption{\small{{\it Left Panel:} The CMB angular power spectrum for $\Lambda$CDM (solid line; $M_{0}^{-1} = 0$, $\nu=1.5$) and general scalaron (dashed line; $M^{-1}_{0}=375 [10^{28}{\rm h^{-1} \hspace{1mm} eV^{-1}}]$, $\nu=1.5$) parameterisation calculated with our modified {\sc camb}. The modified gravity signal is not significant, even over large scales, because the scalaron is too massive. When the quasi-static assumption is relaxed the scalar field affects the larger scales and a noticeable modification to the ISW effect is seen (dotted line; $M^{-1}_{0}= 1.5 \times 10^{4} [10^{28}{\rm h^{-1} \hspace{1mm} eV^{-1}}]$, $\nu=1.5$). {\it Right Panel:} The fractional differences compared to the $\Lambda$CDM cosmology.}}
    \label{fig:2}
  \end{flushleft}
\end{figure*}
\noindent

One might first expect to observe a modified gravity signal in the angular power spectrum on large scales, due to the change in the metric potentials $\phi + \psi$ that modify the ISW effect. However, as we show in Figure~\ref{fig:2} the effect is not particularly significant (solid $\to$ dashed line). The absence of a large modified gravity signal in the low $l$ regime is due to the fact that $f(R)$ models only modify gravity over a particular range of scales. On very large scales the scalaron is too heavy to cause significant deviations from GR, and similarly over very small scales where the density dependent suppression occurs. It is clear from our figures that the ISW effect occurs on scales that are too large to be affected significantly by the scalar field, at least within the reasonable bounds provided by the quasi-static regime. If we consider larger values of $M_{0}^{-1} \gtrsim H_{0}^{-1}$, the scalar field begins to affect the large scales (solid $\to$ dotted line). However this latter plot is for illustrative purposes only, as for these values our approximation scheme and hence field equations are not applicable.

\begin{figure*}
  \begin{flushleft}
    \centering
    \begin{minipage}[c]{1.00\textwidth}
      \centering
      \includegraphics[width=7cm,height=7cm]{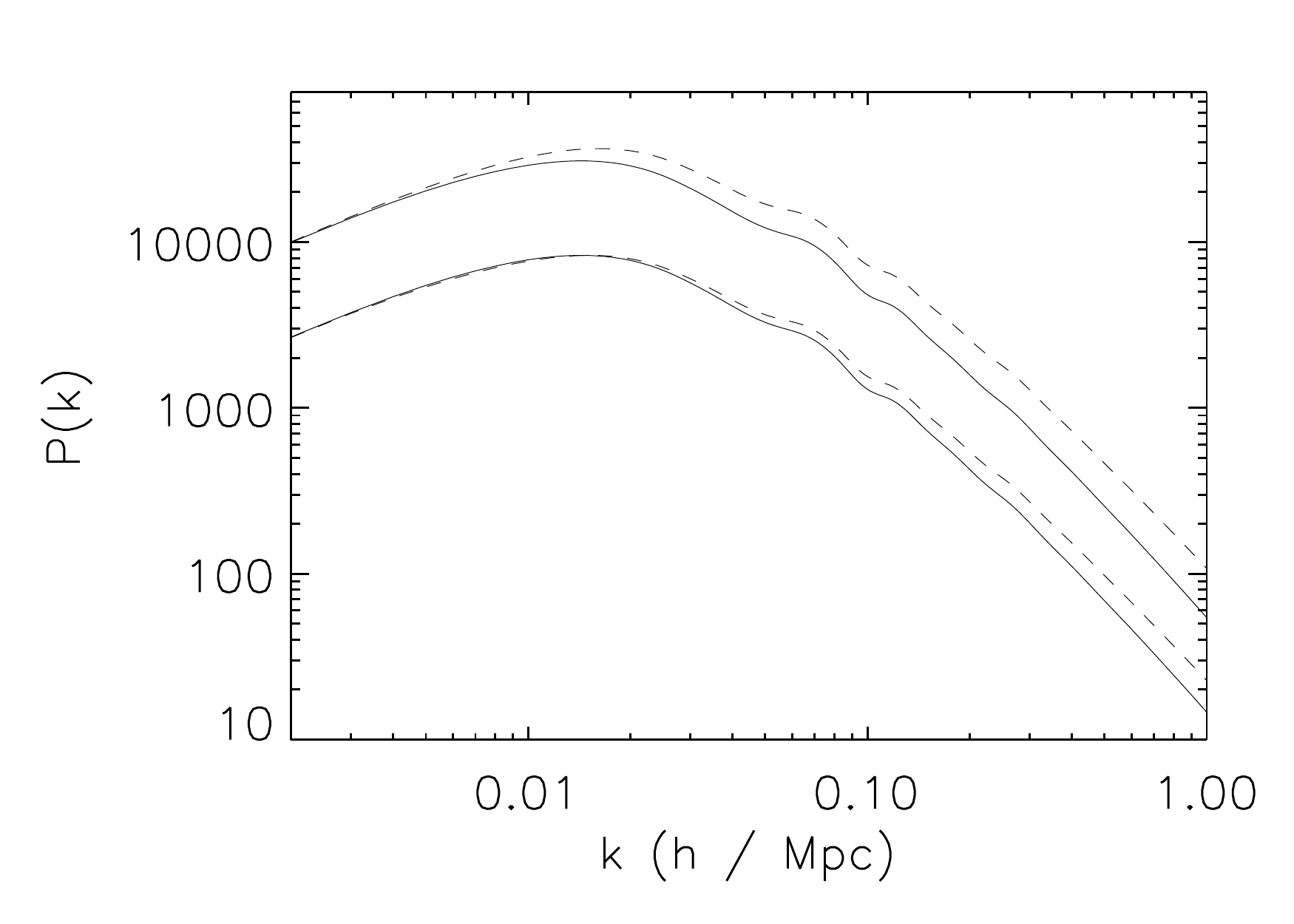}
      \includegraphics[width=7cm,height=7cm]{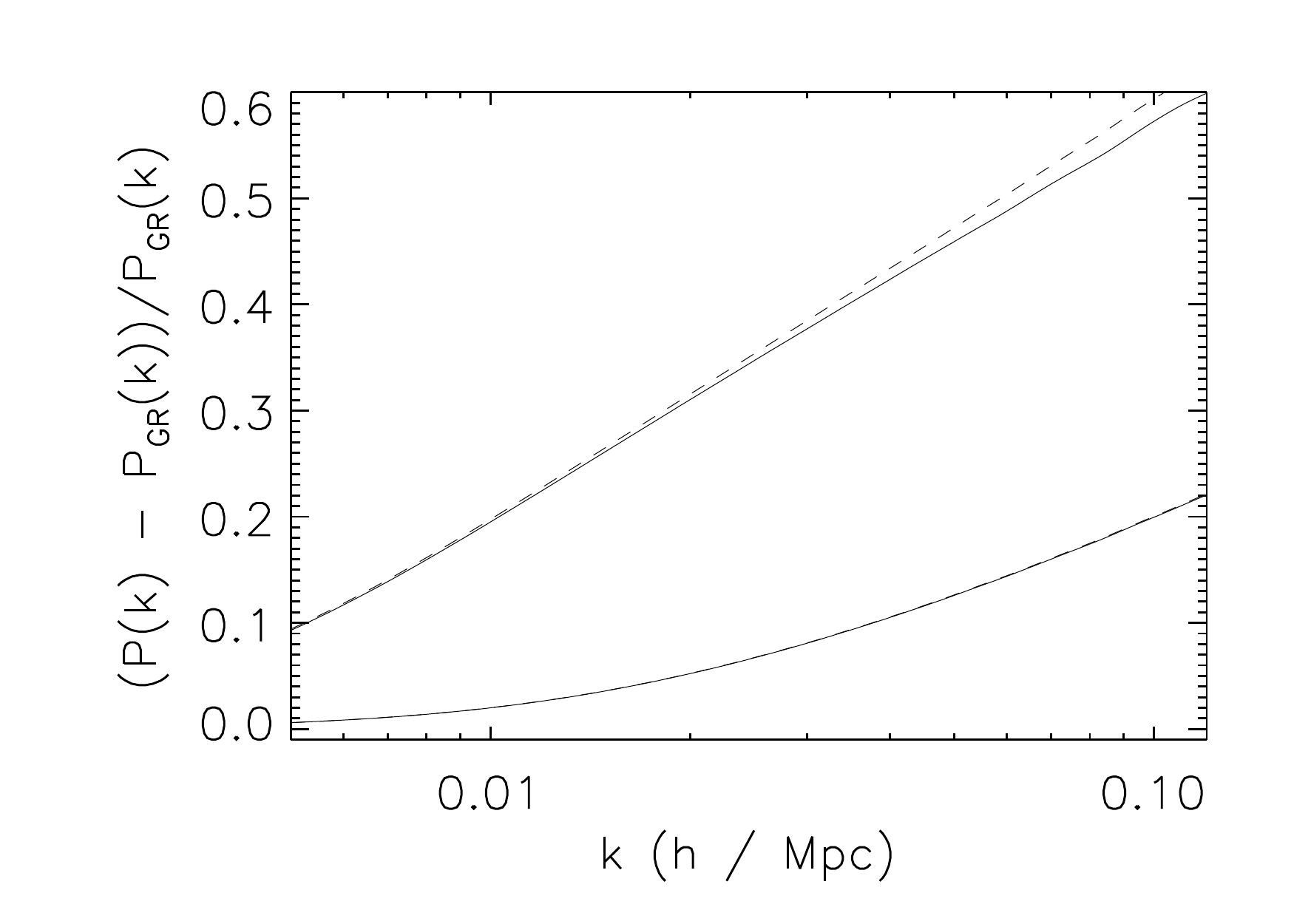}
    \end{minipage}
    \caption{\small{{\it Left Panel:} Linear matter power spectra for $\Lambda$CDM (solid line; $M_{0}^{-1}=0$, $\nu=1.5$) and scalaron (dashed line; $M^{-1}_{0}=375 [10^{28}{\rm h^{-1} \hspace{1mm} eV^{-1}}]$, $\nu=1.5$) cosmologies. The modification to gravity causes a sizeable scale dependent effect in the growth of perturbations. The redshift dependence of the scalaron can be seen by comparing the top and bottom pairs of power spectra evaluated at redshifts $z=0.0$ and $z=1.5$, respectively. {\it Right Panel:} The environmental dependent chameleon mechanism can be seen in the mildly non linear regime. We exhibit the fractional difference $(P(k) - P_{\rm GR}(k))/P_{\rm GR}(k)$ between the f(R) and GR power spectra for the model ($\ref{eq:107}$) with parameters $M^{-1}_{0}=375 [10^{28}{\rm h^{-1} \hspace{1mm} eV^{-1}}]$ and $\nu=1.5$. The dashed lines represent linear power spectra ($P(k)$ and $P_{\rm GR}(k)$ calculated with no higher order effects) and the solid lines are the power spectra calculated to second order. We see that the nonlinearities decrease the modified gravity signal. This is a result of the chameleon mechanism. The top set of lines correspond to $z=0$ and the bottom to $z=0.9$; demonstrating that the modified gravity signal dramatically decreases for larger $z$. This is due to the scalaron mass being much larger at higher redshifts. Furthermore, non linear effects are less significant for increasing $z$. }}
    \label{fig:2b}
  \end{flushleft}
\end{figure*}
\noindent

In addition to the CMB angular power spectrum, we solve ($\ref{eq:p1}$) for $\delta_{\rm m}$ to obtain the linear matter power spectrum. We normalise $P(k,z)$ to its General Relativistic form at $z=100$, where the modified gravity contributions are negligible. Performing these steps and choosing the same parameters as in section \ref{sec:lum} (so as to compare the modified gravity signal between distance and growth based observables), we exhibit the linear matter power spectra at $z=0$ in the left panel of Figure~\ref{fig:2b} (top pair). We now observe a significant effect below a particular lengthscale, as expected. The modification in the power spectrum is also a redshift dependent effect. As $z$ increases the power spectrum returns to its General Relativistic form as a result of the scalaron's mass increasing to the past. This is highlighted by the bottom pair of lines calculated at $z=1.5$ in the same panel.

The lack of signal in the CMB and expansion history compared to the matter power spectrum demonstrates the importance of structure based probes, such as weak lensing, in constraining $f(R)$-type models.

\subsection{Non-linearities} \label{sec:3sub}

There is a vast quantity of useful information in the non-linear regime. However, this sector is one that can bias and invalidate results if not treated carefully. We now briefly discuss approaches to analysing nonlinear physics in the context of modified gravity.

Broadly speaking, we can split the dynamics of the perturbations into three regimes. On large scales one can simply linearise the perturbation equations, where it is not necessary to yet account for non linearities in the fluid equations. Over an `intermediate' range of scales one can consistently treat non-linear effects by calculating the density perturbations to higher order in a perturbative expansion. We first consider this approach, following the recent work \cite{Koyama:2009me} where this problem has been studied in detail. On the smallest of scales however perturbation theory is no longer applicable and one must resort to constructing fitting formulas for the matter power spectrum, calibrated with N-body simulations \cite{Oyaizu:2008sr,Oyaizu:2008tb,Schmidt:2008tn}.

To understand the power spectrum at slightly higher $k$ than the linear theory allows, and to see the effects of the scale dependent chameleon mechanism, we calculate $P(a,k)$ to second order using the {\it closure relations} \cite{Koyama:2009me}. We leave the mathematical details for the interested reader in the Appendix. We solve the associated system of equations for the scalaron model for the two contrasting limits in the gravitational parameters: $M_{0}^{-1}=0.0$, $\nu=1.5$ and $M^{-1}_{0}=375 [10^{28}{\rm h^{-1} \hspace{1mm} eV^{-1}}]$, $\nu=1.5$. The remaining cosmological parameters are set to those used in the previous subsections and throughout the paper. We evolve the relevant equations over the range $k = (0.001,0.13) {\rm h Mpc^{-1}}$, however the perturbative solution will approximately break down before $k = 0.13$ ${\rm h Mpc}^{-1}$. At precisely what scale the second order solution ceases to be accurate is not clear and will be model and redshift-dependent. However, we expect it to be of order $k_{\rm nonlinear} \sim 0.1$ ${\rm h Mpc}^{-1}$ at $z=0$.

In the right panel of Figure~\ref{fig:2b} we illustrate the modified gravity signal in the matter power spectrum; specifically we plot the fractional difference between the $f(R)$ and GR power spectra $(P(k) - P_{\rm GR}(k))/P_{\rm GR}(k)$ at redshift $z=0$ (top pair) and $z=0.9$ (bottom pair). The dashed lines represent both $P(k)$ and $P_{\rm GR}(k)$ in the linear regime, and the solid lines represent both using the second order calculation described in the Appendix. We observe that  the nonlinear terms reduce the modified gravity signal, and have a tendency to return the power spectrum back to its General Relativistic form. This effect is essentially due to the backreaction of the density perturbations on the mass of the scalar field; the scalar field's mass is not just sensitive to the background energy density, but also the perturbations when one considers nonlinear terms in the fluid equations. This behaviour will not in general be captured in current linear to non linear mappings such as {\sc halofit} \cite{Smith03}.

The previous Figure shows that nonlinear effects for these models are extremely small in the regime $0.001$ ${\rm h Mpc}^{-1} < k \lesssim 0.1$ ${\rm h Mpc}^{-1}$ and at redshift $z = 0.9$; they constitute an order $\lesssim {\cal O}(2\%)$ effect to the power spectrum with the parameter choices $M^{-1}_{0}=375 [10^{28}{\rm h^{-1} \hspace{1mm} eV^{-1}}]$, $\nu=1.5$. In section \ref{sec:5} we consider potential constraints on these models from future weak lensing surveys, and the above calculation allows us to consistently use the linear matter power spectrum on scales $k \lesssim 0.1$ ${\rm h Mpc}^{-1}$. 

Although nonlinear effects are relatively small for the ranges under consideration, and although the perturbative method does not propagate far into $k$, perturbation theory is not without use as it can be used to calibrate or test methods used in the fully non linear regime. One approach to obtaining the power spectrum on scales $k \gtrsim 0.1 {\rm h Mpc}^{-1}$ is to use the Parameterized Post Friedmannian (PPF) fitting formula for $f(R)$ models, developed in \cite{Hu:2007pj}

\begin{equation} \label{eq:ap10} P(a,k) = { P_{\rm no-cham}(a,k) + c_{\rm nl} \Sigma^{2}(a,k) P_{\rm GR}(a,k) \over 1 + c_{\rm nl} \Sigma^{2}(a,k)}    \end{equation}

\noindent where $P_{\rm no-cham}$ is the power spectrum obtained if one were to assume that General Relativity is modified on all scales (that is, neglecting the nonlinear effects which drive the power spectrum back to its General Relativistic form). $c_{\rm nl}$ and $\Sigma^{2}(a,k)$ are functions that can be obtained by comparing full non linear N-body results to the semi-analytic calculation of $P(a,k)$ in the mildly non linear regime.

As shown in \cite{Koyama:2009me}, the fitting function ($\ref{eq:ap10}$) is extremely accurate on scales $k \lesssim 1{\rm h Mpc}^{-1}$, but only if we have knowledge of $P_{\rm no-cham}$ in the nonlinear regime from simulations. Whilst promising work has been made in this direction \cite{Zhao10,Oyaizu:2008sr,Oyaizu:2008tb,Schmidt:2008tn}, there are currently insufficient N-body results to calibrate the fitting function for different scalaron parameters. It is for this reason we do not use ($\ref{eq:ap10}$) in the next section of the paper, as it does not guarantee a better fit to $f(R)$ models than {\sc halofit} in the absence of an accurate description of $P_{\rm no-cham}$. Moreover, an industrious programme of N-body simulations to explore the gravitational parameter space is unlikely to be embarked upon unless a suitable, general and physical parameterisation/framework is settled upon. Addressing this is one of the intentions of this work.

\section{Weak Lensing} \label{sec:4}

Weak lensing is a statistical measure of the distortion of distant source galaxies. It occurs due to gravitational shear induced by the intervening mass distribution, and so observing the weak lensing signal yields information on the distribution of matter projected along the line of sight. 

With respect to analysing cosmological models, of particular interest is the convergence $\kappa$. This is related to the shear matrix $\Phi_{ij}$ via the relation

\begin{equation} \kappa = {\Phi_{11} + \Phi_{22} \over 2} \end{equation} 

\noindent where the shear matrix is in turn related to the Newtonian potential $\phi$ as

\begin{equation} \Phi_{ij} = \int^{\chi_{h}}_{0} d\chi g(\chi) \partial_{i}\partial_{j}\phi \end{equation} 

\noindent and 

\begin{equation} g(\chi) = 2\int^{\chi_{h}}_{\chi} d\chi' n(\chi') { r(\chi)r(\chi'-\chi) \over r(\chi')} .\end{equation}

\noindent where $r(\chi)$ is the comoving angular diameter distance and $n(\chi)$ is the probability of finding a galaxy at comoving distance $\chi$. The resulting convergence power spectrum is given by \cite{Bartelmann01}
\begin{equation} \label{eq:81} P_{\kappa}(l) = \frac{9 \Omega_{m}^{2} H_{0}^{2}}{4c^{4}} \int^{\chi_{h}}_{0}d\chi \left[{g(\chi) \over a r(\chi)}\right]^{2} P\left({l \over r},\chi\right) \end{equation} 

\noindent where $P(k,\chi)$ is the matter power spectrum and $\chi_{h}$ is the comoving distance to the horizon. 

It is clear therefore that weak lensing is sensitive to the underlying matter power spectrum and the expansion history. This makes the probe highly capable of distinguishing between modified gravity and dark energy models. In addition, a significant advantage to using weak lensing is that the signal is sourced by the metric potentials and is therefore not subject to any unknown galaxy bias. Modified gravity models can also act to alter the relation between the power spectrum of the metric potentials $P_{\phi \psi}$ and the matter power spectrum $P_{\delta}$ \cite{Thomas08,Amendola:2007rr}, which is expressed finally in (\ref{eq:81}). 

Weak lensing does face technical challenges in the form of stringent requirements for shape estimation \cite{Kitching10} and intrinsic alignments \cite{Joachimi10} but the potential power of this method for dark energy, neutrino masses and testing gravity provides vast motivation for solving them. We focus on this probe of the matter power spectrum in the next section and combine it with the CMB after.

\section{Future Surveys and their Gravitational Limits} \label{sec:5}

\subsection{Weak Lensing: Euclid}

We perform a Fisher matrix analysis to calculate expected parameter sensitivity for a future weak lensing survey: the proposed space-based Euclid mission \cite{Refregier10} (see \cite{Narikawa:2009ux,Martinelli10} for related works). While one can undertake a more studious MCMC analysis for this purpose, as in \cite{Martinelli10}, the Fisher matrix method produces bounds that are essentially equivalent for these illustrative purposes \cite{Martinelli10}. Euclid intends to measure galaxies over a sky coverage of $20'000$ square degrees with an effective galaxy density of $40$ $\mathrm{gal/arcmin^{2}}$. The corresponding Fisher matrix is given by

\begin{equation} F_{ij} = \sum_{l} {\partial C \over \partial p_{i}}{\rm Cov}^{-1} {\partial C \over \partial p_{j}}   \end{equation} 

\noindent where $C$ is the weak lensing observable

\begin{equation} C_{ij}(l) = P_{ij} + \langle \gamma_{\rm int}^{2} \rangle {\delta_{ij} \over \bar{n}_{\rm i} }.\end{equation}

\noindent $P_{ij}$ is given in (\ref{eq:81}) with indices denoting tomographic redshift bins, and $\bar{n}_{i}$ is the average galaxy number per steradian in bin $i$. The quantity $\langle \gamma_{\rm int}^{2} \rangle^{1/2}$ is the rms intrinsic shear in each component, which we set to $0.22$. $\partial C / \partial p_{\rm i}$ is the derivative of the weak lensing observable with respect to the parameter $p$ being varied. ${\rm Cov}$ is the covariance matrix

\begin{equation} {\rm Cov} \left[ C_{ij}^{k}(l),C_{kl}^{k} \right] = {\delta_{\rm ll'} \over (2l + 1)f_{\rm sky} \triangle l} \left[ C_{ik}^{k}(l)C_{jl}^{k}(l) + C_{il}^{k}(l)C_{jk}^{k} \right] . \end{equation}

\noindent and $f_{\rm sky}$ is the sky fraction. To model the redshift distribution we use the expression

\begin{equation} n(z) \propto z^{\alpha} {\rm exp}[-(z/z_{0})^{\beta}] \end{equation}

\noindent with $z_{0} = z_{\rm m}/1.412$, $\alpha = 2$, $\beta = 1.5$. $z_{\rm m}$ is the median redshift and for a Euclid-like survey $z_{\rm m} = 0.9$ \cite{Refregier10,Cimatti:2009is}. We consider five tomographic redshift bins, chosen such that there are approximately an equal number of galaxies in each. Finally, to account for the photometric error we convolve the redshift bins with a Gaussian of width $\sigma_{z} = \sigma_{p}(1+z)$, where $\sigma_{p} = 0.03$.

\subsection{The CMB: Planck}

In addition to the weak lensing survey, we use a forecast for the upcoming Planck surveyor \cite{:2006uk}. Similarly the CMB Fisher matrix is given by
\begin{equation}
  F_{ij}^{CMB}=\sum_{l}\sum_{X,Y}\frac{\partial
    C_{X,l}}{\partial p_{i}}\mathrm{COV^{-1}_{XY}}\frac{\partial
    C_{Y,l}}{\partial p_{j}},
  \label{eqn:cmbfisher}
\end{equation} where $C_{X,l}$ is the harmonic power spectrum for the temperature--temperature ($X\equiv TT$), temperature--E-polarization ($X\equiv TE$) and the E-polarization--E-polarization ($X\equiv EE$) power spectra. The covariance matrix $\rm{COV}_{XY}$ for the power spectra is given, under a Gaussian assumption, by, 
\begin{eqnarray}
{\rm COV}_{T,T} & = & f_\ell\left(C_{T,l}+W_T^{-1}B_l^{-2}\right)^2 \\
{\rm COV}_{E,E} & = & f_\ell\left(C_{E,l}+W_P^{-1}B_l^{-2}\right)^2  \\
{\rm COV}_{TE,TE} & = & f_\ell\Big[C_{TE,l}^2+\left(C_{T,l}+W_T^{-1}B_l^{-2}\right)\left(C_{E,l}+W_P^{-1}B_l^{-2}\right)\Big]  \\
{\rm COV}_{T,E} & = & f_\ell C_{TE,l}^2  \\
{\rm COV}_{T,TE} & = & f_\ell C_{TE,l}\left(C_{T,l}+W_T^{-1}B_l^{-2}\right) \\
{\rm COV}_{E,TE} & = & f_\ell C_{TE,l}\left(C_{E,l}+W_P^{-1}B_l^{-2}\right)\; ,
\end{eqnarray}
where $f_\ell = \frac{\ell}{(2\ell+1)f_{\rm sky}}$ and $W_{T,P}=(\sigma_{T,P}\theta_{\rm fwhm})^{-2}$  is the weight per solid angle for temperature and polarization, with a $1\sigma$ sensitivity per pixel of $\sigma_{T,P}$ with a beam width of $\theta_{\rm fwhm}$. The beam window function is expressed in terms of the beam full width half maximum as $B_\ell = \exp\left(-\ell(\ell+1)\theta_{\rm fwhm}^2/16\ln2\right)$. 

In the forecast we act conservatively and use only one of Planck's channels (143 GHz) for the signal assuming the other frequencies are used in foreground removal. The channel has a beam of $\theta_{\rm fwhm}=7.1'$ and sensitivities of $\sigma_T = 2.2 \mu K/K$ and $\sigma_P = 4.2\mu K/K$. We take the sky fraction to be $f_{\rm sky} = 0.80$ due to Galactic coverage. 

Over the lowest multipoles one needs an accurate treatment of the Integrated Sachs-Wolfe (ISW) effect via perturbations. Due to uncertainty in this treatment and potential polarisation foreground contamination we exclude the lowest multipoles from our analysis. Instead we utilise the range $30 \le \ell \le 2000$. A consequence of this is slightly weaker limits owing to a poorer determination of the optical depth to reionisation $\tau$. However, we feel that conservativeness is preferable to an incorrect or overly optimistic treatment. This cut in scale also has limited affect on the direct determination of the scalaron parameters. This is because, as we demonstrated earlier, the CMB is largely insensitive to gravitational modifications of this kind. This was shown in Figure~\ref{fig:2}. Therefore, in this analysis the role of the Planck fisher matrix will be to act as a {\it prior} on each of the main cosmological parameters and aid the limits by breaking parameter degeneracies.

For the inclusion of the Planck Fisher matrix we {\it start} from the fiducial parameter set $\Omega_{m} h^{2}$, $\theta_S$, $\mathrm{ln} (A_S)$, $\Omega_{b} h^{2}$, $n_{s}$ and $\tau$, where $\theta_S$ is the angular size of the sound horizon at last scattering and $\mathrm{ln} (A_S)$ is the logarithm of the primordial amplitude of scalar perturbations. This is a more Gaussian parameter set as detailed by the Dark Energy Task Force (DETF)\cite{Albrecht:2006um}. We calculate the Planck CMB Fisher matrix using {\sc camb} for our power spectra. The optical depth acts as a nuisance parameter for the matter power spectrum and so we marginalise over it in our analysis. We {\it then} transform the resulting parameters to Euclid's basis set: $\Omega_{m}$, $h$, $\mathrm{ln} (A_S)$, $\Omega_{b}$ and $n_{s}$. The resulting Fisher matrix can then be added to the weak lensing matrix. The transformation is performed using the Jacobian
\begin{equation}
J_{\hat {\beta} \beta} = \frac{\partial
  p_{\beta}}{\partial \hat{p}_{\hat{\beta}}}
\end{equation}
and the final basis of the Fisher matrix is determined by
\begin{equation}
\hat{\mathbf{F}} = {\mathbf J}{\mathbf F}{\mathbf J}^T\;.
\end{equation}

\begin{figure*}[!h]
  \begin{flushleft}
    \centering
    \begin{minipage}[c]{1.00\textwidth}
      \centering
      \includegraphics[width=7cm,height=7cm]{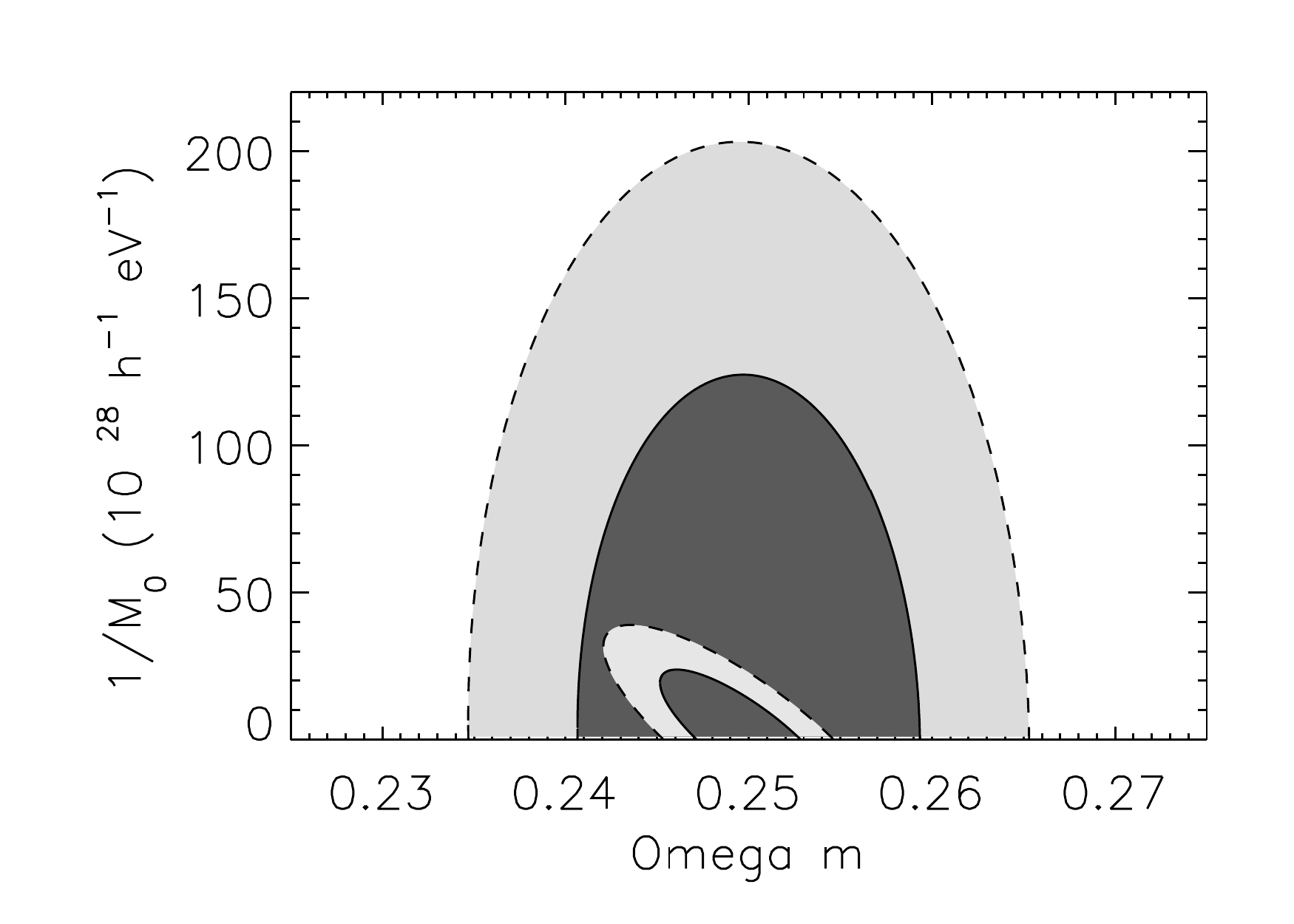}
      \includegraphics[width=7cm,height=7cm]{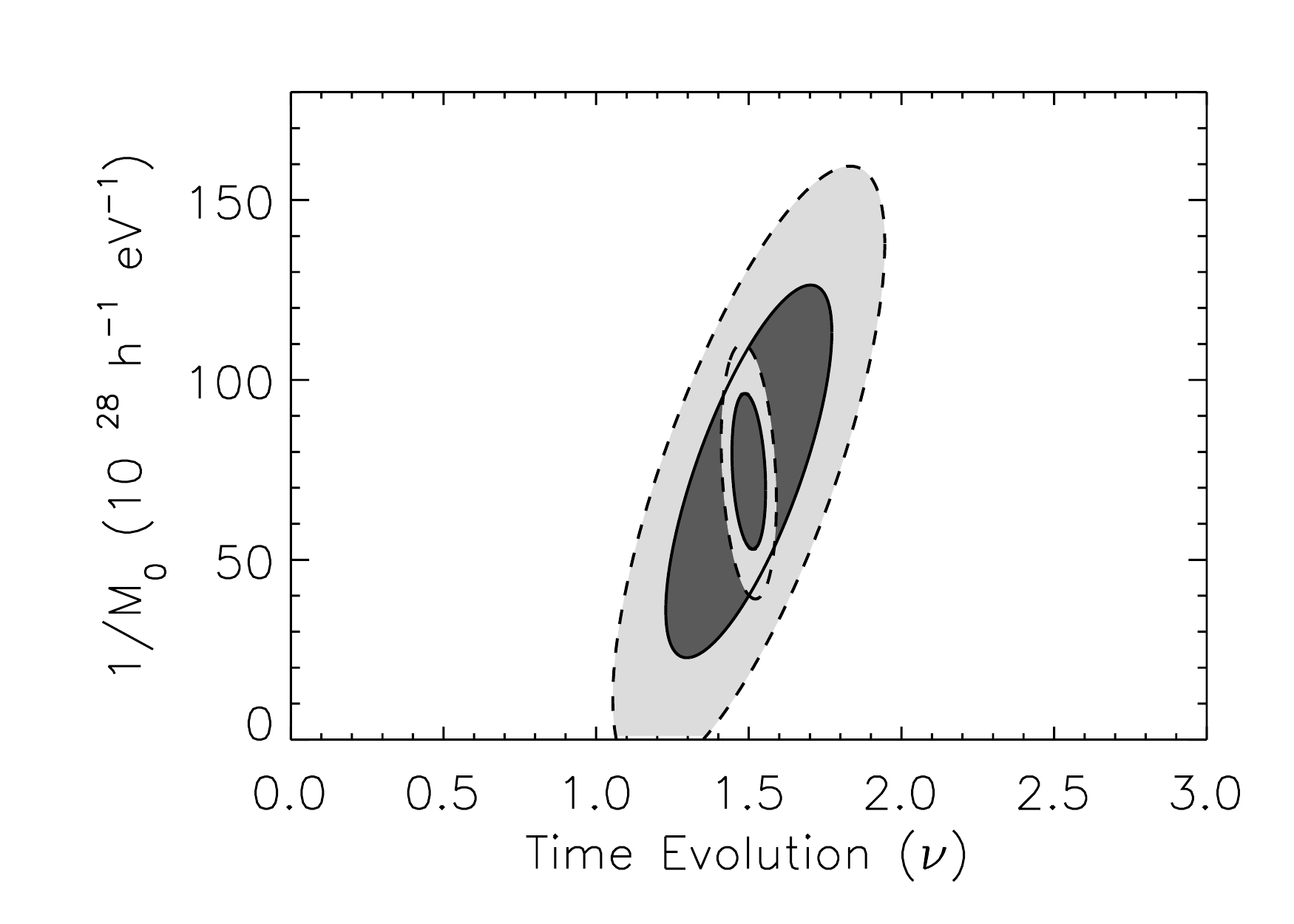}\\
      \includegraphics[width=7cm,height=7cm]{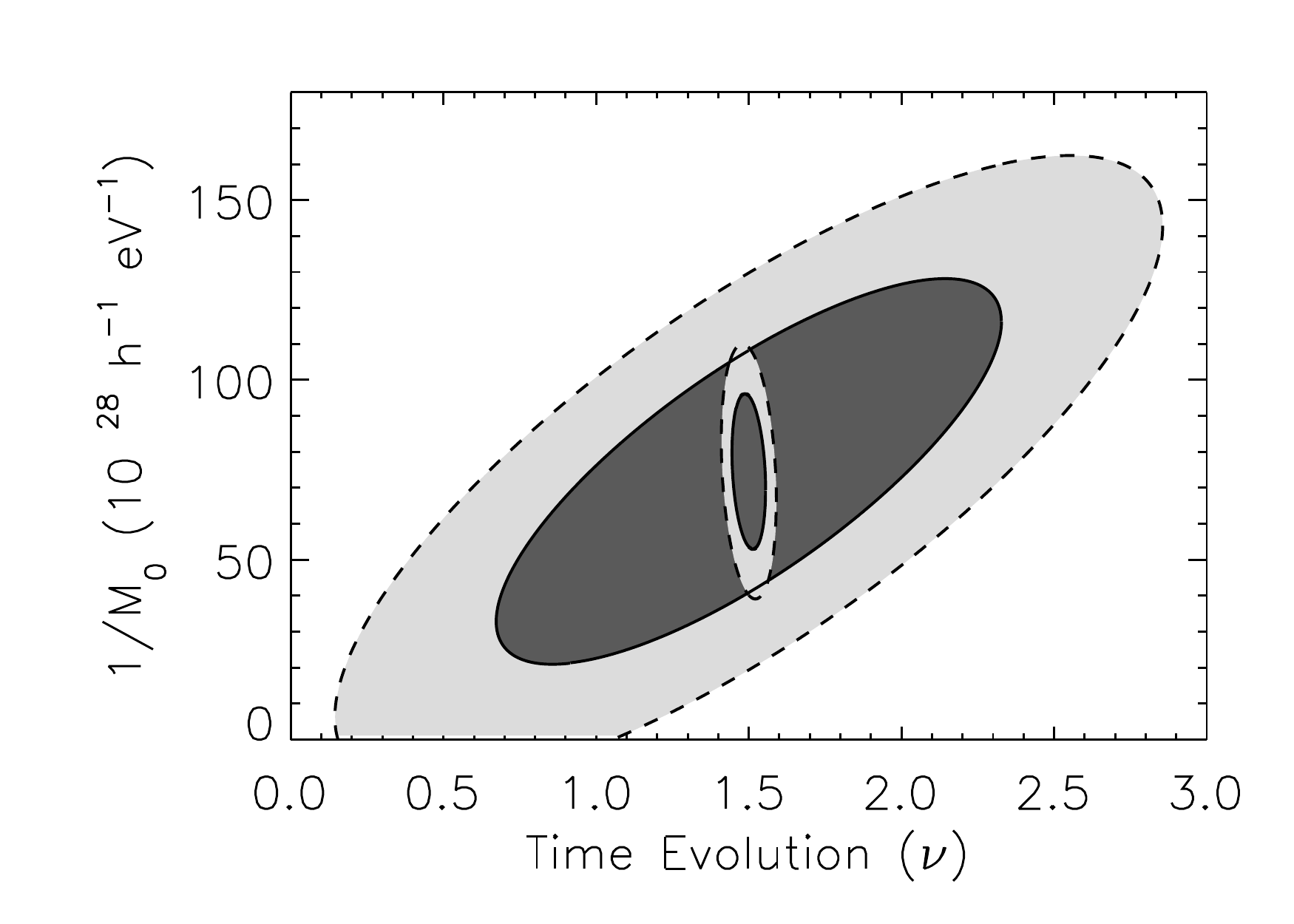}
      \includegraphics[width=7cm,height=7cm]{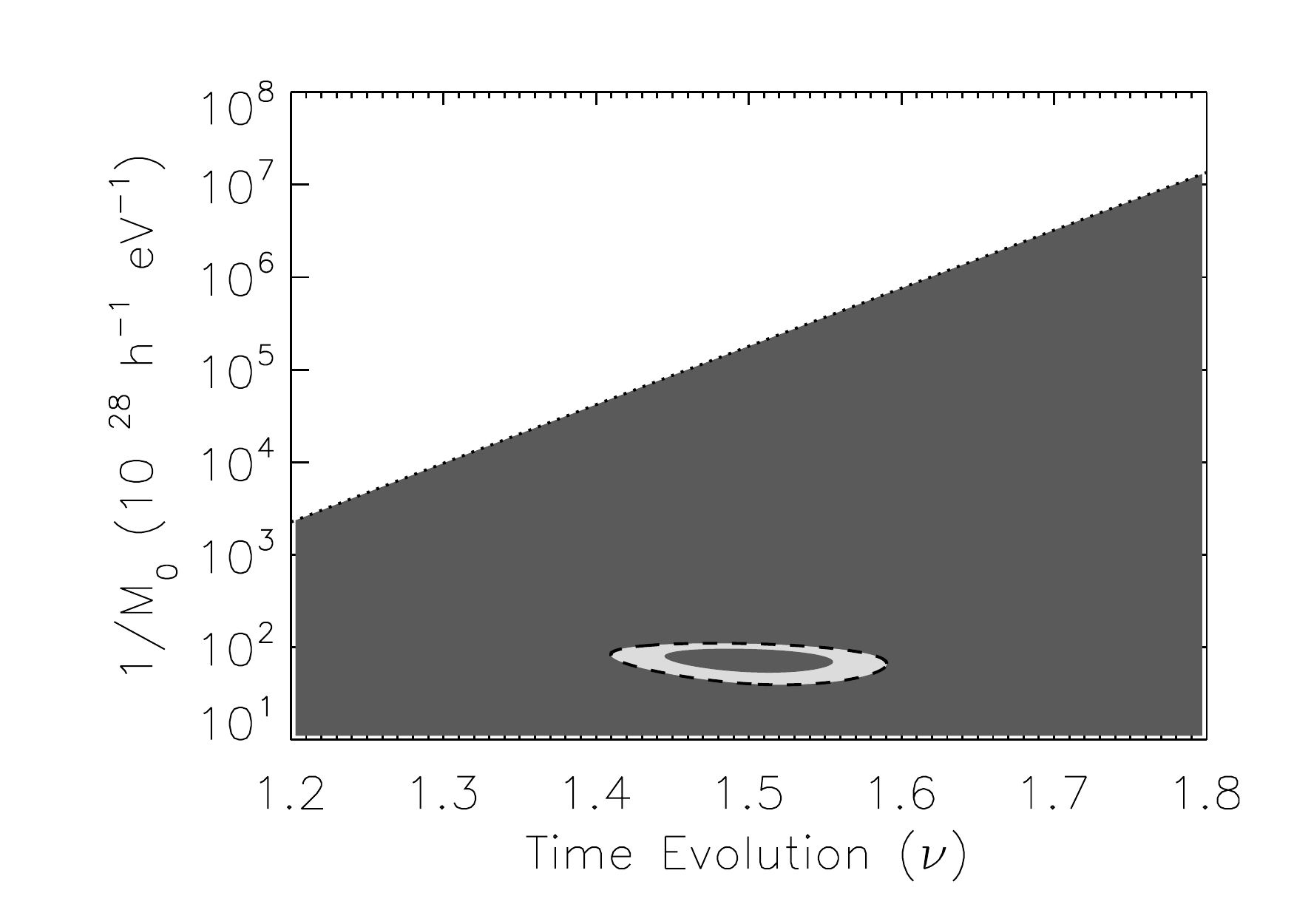}

    \end{minipage}
    \caption{\small{{\it Top Left Panel:} Euclid-only projected bounds on the scalaron with fiducial values $M_{0}^{-1}=0$ ($\Lambda$CDM) and $\nu=1.5$ (fixed). The wider set of limits are for linear scales ($\ell_{\mathrm{max}}=400$) and the tighter set are with all scales included ($\ell_{\mathrm{max}}=10,000$). In {\it all} plots solid (dashed) lines represent $68\%$ ($95\%$) confidence limits. {\it Top Right Panel:} Similarly the improvement from using non linear data is evident when both parameters are varied ($M^{-1}_{0}=75 [10^{28}{\rm h^{-1} \hspace{1mm} eV^{-1}}]$, $\nu=1.5$ fiducial): $\ell_{\mathrm{max}}=400$ (less constraining) $\to$ $\ell_{\mathrm{max}}=10,000$ (most constraining). It is not possible to currently treat this regime accurately but the improvement is indicative of the gain possible. Both sets of contours are for Euclid $+$ Planck. {\it Bottom Left Panel:} The specific improvement from including Planck (tighter set) can be seen when added to Euclid-only full non linear limits (wider set). {\it Bottom Right Panel:} The combined limit from the previous panel is shown in contrast to current solar system constraints (dotted line; region above excluded). The models are successful in evading such tests (note the log vertical axis) and so weak lensing surveys will provide essential and far superior bounds.}}
    \label{fig:3}
  \end{flushleft}
\end{figure*}
\noindent

\subsection{Forecast Results and Analysis}

For the fiducial model, we take the same parameters used throughout the paper: $\Omega_{m0} = 0.25$, $\Omega_{b0} = 0.05$, $h=0.7$, $\mathrm{ln} (A_S) = \ln (2.34 \times 10^{-9})$ and $n_{\rm s} = 1$. The equation of state is effectively fixed to $w = -1$ and $w_{a} = 0$ given our earlier arguments in Section~\ref{sec:lum} regarding the expansion history. We make the simplifying assumptions of zero spatial curvature $\Omega_{k0} = 0$ and that there are no massive neutrinos. We note that including massive neutrinos could act to introduce an additional degeneracy, as they suppress the growth of structure \cite{Motohashi:2010sj}.

The $\Lambda$CDM limit of the scalaron parameterisation in ($\ref{eq:107}$) corresponds to $M_{0}^{-1} \to 0$. We wish to use this cosmology as one of the fiducial models, however if we take $M_{0}^{-1}=0$ the resulting observables are completely insensitive to $\nu$ due to there being no modified gravity signal. In this case the fisher matrix is non-invertible and we will be unable to derive forecast constraints on the cosmological parameters. It follows that when we take $\Lambda$CDM as the fiducial model, we must temporarily fix the time evolution: $\nu=1.5$. Another procedure is to choose $M_{0}^{-1} \neq 0$ and vary both modified gravity parameters. We therefore also undertake analyses with fiducial cosmologies in a modified framework: $M^{-1}_{0}=75 [10^{28}{\rm h^{-1} \hspace{1mm} eV^{-1}}]$ and $\nu=1.5$.

We perform our analyses in two regimes: in the first instance we only consider weak lensing contributions from linear scales $10 < l < 400$. The resulting constraints that we obtain will be conservative but accurate, as we have a firm understanding of the perturbations in this range. For the second set of analyses we include the full range of scales $10<l<10000$, however in doing so we include non linear effects. In this work we use the {\sc halofit} fitting formula to reconstruct the non linear power spectrum. We argued earlier that with the present limited number of simulations this was at least as good as the new formalism given by ($\ref{eq:ap10}$). We note that this approach will not yield an accurate description of the nonlinear regime for these modified gravity models at present, however the resulting forecasts will be indicative of the gain that {\it could} be obtained if we had a suitable description of the power spectrum on small scales.

The limits on these models are illustrated in Figure~\ref{fig:3} for Euclid only and Euclid and Planck combined. We have also included the constraints arising from solar system tests of these models. A discussion of the solar system bound is included in the Appendix.

It is clear that weak lensing will be most adept at probing the nature of gravity: The Euclid probe is capable of stringent constraints on the general gravitational parameterisation introduced and examined in this work. However, Figure~\ref{fig:3} highlights the significance in understanding the non linear regime in this gravitational framework. The additional information, particularly where the scale dependent signal is evident, leads to vastly reduced bounds and could be vital in making a definite measurement of new gravity. This is exhibited clearly in the top right panel. 

The bottom left panel illustrates the great importance of complementarity in cosmology. Despite the fact that Euclid is essential in providing the actual modified gravity signal, Planck is highly beneficial in reducing the bounds. This contribution is an indirect one and it occurs by reducing the parameter space. Specifically, Planck tells us the primordial amplitude $A_{s}$ and spectral index $n_{s}$ of the $\sim$ unmodified power spectrum (Figure~\ref{fig:2}) with exquisite detail. Therefore any subsequent change to the shape or, approximately, the amplitude of the matter power spectrum, as in Figure~\ref{fig:2b}, must be the result of alterations in gravity through $\mu$ and $\nu$. Indeed, this is exactly what we see; Planck breaks the degeneracy mainly through the spectral index and, to a slightly lesser extent, the amplitude. This explains the much tighter combined limit in the bottom left panel of Figure~\ref{fig:3}. 

The overall degeneracy between the modified gravity parameters for lensing only can simply be understood by considering that as the scalaron's mass decreases there is a larger modified gravity signal and a boost to the power spectrum. To compensate this and to give the same observable one can increase the time evolution through $\nu$ such that the scalaron decays back to GR faster in the recent past. Finally, the bottom right panel in Figure~\ref{fig:3} demonstrates the absolute necessity in using cosmological probes to explore gravity and this general parametersation. The density dependent suppression makes it severely challenging even for precision solar system tests.

\section{Conclusions} \label{sec:6}

We have examined a general scheme for a class of viable gravitational $f(R)$ models; the scalaron framework. This is physically motivated and gives rise to distinctive observational features. For viable $f(R)$ models the background expansion is nearly identical to $\Lambda$CDM, a fact reflected in a luminosity distance-redshift relationship that is practically indistinguishable from the standard cosmology. By solving the modified perturbation equations with an augmented {\sc camb}, we also find that the CMB does not add information on the modified signal directly, because the scalaron mass is too large to affect the ISW substantially. The most significant observational effect for these models occurs in the matter power spectrum; specifically $P(a,k)$ acquires a redshift and scale dependent modification relative to GR. This signal can be used to constrain modified gravity parameters with weak lensing.

By examining parameterisations of $f(R)$ gravity in detail we have argued that existing functional forms in the literature do not capture the essential features of $f(R)$ models at late times $z \lesssim 1$.  To accurately describe the behaviour of these models for $z \lesssim 1$ we have directly parameterized the scalaron mass $M(a)$. Whilst our approach can be related to the standard $\mu(a,k)$ and $\eta(a,k)$ parameterization in the literature \cite{Martinelli10,Bertschinger08,Zhao:2008bn,Daniel10} (and also the $\gamma$ parameterization; see \cite{Appleby:2010dx,Motohashi:2010tb,Fu:2010td}), the mass of the scalar field is an ideal function to parameterize, in the sense that it is independent of scale, only weakly dependent on cosmological parameters and has a clear physical interpretation.

Using the general scalaron model ($\ref{eq:107}$), we found that a future weak lensing probe such as Euclid will be able to place extremely tight constraints on deviations to gravity. We quote our final results as  $M_{0} = 1.34 \pm 0.62 \times 10^{-30} [{\rm h \hspace{1mm} eV}] $, $\nu = 1.50 \pm 0.18$ for $l < 400$ and $M_{0} = 1.34 \pm 0.25 \times 10^{-30}  [{\rm h \hspace{1mm} eV}]  $, $\nu = 1.50 \pm 0.04$ for the $l < 10000$ (Planck + Euclid) combined analysis. Our forecast results highlight a number of interesting points, specifically; the importance of constraining these models with cosmological weak lensing data (compared to solar system tests), the value of using CMB data to break parameter degeneracies, and also the importance of modeling the nonlinear power spectrum. As stated previously, this regime is particularly hard to treat in any gravitational theory, and could bias the results if implemented blindly.  While some approaches, such as higher order perturbation theory, can give us an indication of physics at nonlinear scales, it is evident that there is a huge demand for programmes of simulations and a better approach to fitting to them. A detailed study of this problem will be presented elsewhere.

\vspace{5mm}

\section*{Acknowledgments} 
{\em Acknowledgments: The work of SA and JW was supported by the DFG TRR 33 `The Dark Universe'. SA would like to thank Scott Daniel and Eric Linder for discussions that have greatly assisted in modifying CAMB.}

\section{\label{sec:a1}Appendix I: Nonlinear regime}

The mildly nonlinear regime can be treated semi-analytically using a standard, order by order expansion of the perturbation equations. Specifically, for $f(R)$ models one can write the fluid equations as \cite{Koyama:2009me} 

\begin{eqnarray}  \label{eq:ap1} & & {\partial \Phi_{a} \over \partial \tau} + \Omega_{ab}(k,\tau) \Phi_{b} =  \int {d^{3}{\bf k}_{1} d^{3}{\bf k}_{2} \over (2\pi)^{3}} \delta_{\rm D}({\bf k} - {\bf k}_{1}- {\bf k}_{2}) \gamma_{abc}({\bf k}_{1}, {\bf k}_{2},\tau) \Phi_{b}({\bf k}_{1},\tau) \Phi_{c}({\bf k}_{2},\tau) \\ \nonumber & & + \int {d^{3}{\bf k}_{1} d^{3}{\bf k}_{2} d^{3}{\bf k}_{3}  \over (2\pi)^{6}} \delta_{\rm D}({\bf k} - {\bf k}_{1}- {\bf k}_{2}- {\bf k}_{3}) \sigma_{abcd}({\bf k}_{1}, {\bf k}_{2},{\bf k}_{3}, \tau) \Phi_{b}({\bf k}_{1},\tau) \Phi_{c}({\bf k}_{2},\tau) \Phi_{d}({\bf k}_{3},\tau) \end{eqnarray}

\noindent where we have used the linearized Einstein equations to remove the metric potentials, and have defined

\begin{equation} \Phi_{a} = \left( \begin{array}{c}
\delta(\tau, {\bf k})\\
-\theta(\tau,{\bf k}) \end{array} \right) \end{equation}

\begin{equation} \Omega_{ab} = \left( \begin{array}{cc}
0 & -1  \\
-{4\pi G \rho_{\rm m} \over H^{2}}\left( 1 + {k^{2} \over 3a^{2}\Pi(a,{\bf k})}\right) & 2+ {\dot{H} \over H^{2}}  \end{array} \right) \end{equation}

\begin{eqnarray} & & \gamma_{112} ({\bf k}_{1}, {\bf k}_{2},\tau)  = {1 \over 2} \left(  1 + {{\bf k}_{1}.{\bf k}_{2} \over |{\bf k}_{1}|^{2}}\right) \\ & &  \gamma_{121}({\bf k}_{1}, {\bf k}_{2},\tau)  = {1 \over 2} \left(  1 + {{\bf k}_{1}.{\bf k}_{2} \over |{\bf k}_{2}|^{2}}\right) \\ & &    \gamma_{222}({\bf k}_{1}, {\bf k}_{2},\tau)  = {1 \over 2}{{\bf k}_{1}.{\bf k}_{2} |{\bf k}_{1} +{\bf k}_{2}|^{2} \over |{\bf k}_{1}|^{2} |{\bf k}_{2} |^{2}}          \\ & & \gamma_{211}({\bf k}_{1}, {\bf k}_{2},\tau)  = -{1 \over 12H^{2}}\left( {8\pi G \rho_{\rm m} \over 3}\right)^{2} {(k_{1} + k_{2})^{2} \over a^{2}} {M_{2}(a) \over \Pi(\tau,{\bf k}_{12}) \Pi(\tau,{\bf k}_{1}) \Pi(\tau,{\bf k}_{2})} \end{eqnarray}

\begin{eqnarray} \nonumber & &  \sigma_{2111}({\bf k}_{1},{\bf k}_{2},{\bf k}_{3},\tau) = -{1 \over 36H^{2}} \left({8\pi G \rho_{\rm m} \over 3}\right)^{3} {k_{123}^{2} \over a^{2}} {M_{3}(a) \over \Pi({\bf k}_{123},a)\Pi({\bf k}_{1},a)\Pi({\bf k}_{2},a)\Pi({\bf k}_{3},a)} \\ & & \times \left( 1 - {1 \over 3 M_{3}(a)} \left[ {(M_{2}(a))^{2} \over \Pi({\bf k}_{23})} + {\rm permutations}\right] \right) \end{eqnarray}

\begin{equation} \Pi(a,{\bf k}) = \left( {k^{2} \over a^{2}} + {M^{2} \over 3} \right)  \end{equation}

\begin{eqnarray} & & M_{2}(a) = 3{d M^{2} \over df_{R}} \\ & & M_{3} = {d M_{2} \over df_{R}} \end{eqnarray}

\noindent and $\tau = \log[a]$. By using the standard expansions $\Phi_{a} = \Phi_{a}^{(1)} + \Phi_{a}^{(2)} + \Phi_{a}^{(3)} + ...$ one can solve ($\ref{eq:ap1}$) and obtain the power spectrum to any order using the definition

\begin{equation} (2\pi)^{3} \delta_{D}({\bf k} + {\bf k}') P_{ab}(|{\bf k}|,\tau) = \langle \Phi_{a}({\bf k},\tau), \Phi_{b}({\bf k}',\tau) \rangle \end{equation}

However, rather than using this approach we instead follow \cite{Koyama:2009me} and directly calculate the matter power spectrum to second order, using the so-called closure relations. Specifically one defines two additional statistical quantities in addition to $P_{ab}(a,k)$;

\begin{eqnarray} & & (2\pi)^{3} \delta_{D}({\bf k} + {\bf k}') R_{ab}(|{\bf k}|,\tau,\tau') = \langle \Phi_{a}({\bf k},\tau), \Phi_{b}({\bf k}',\tau') \rangle   \qquad \tau > \tau'   \\ & & (2\pi)^{3} \delta_{D}({\bf k} - {\bf k}') G_{ab}(|{\bf k}|,\tau,\tau') = \left \langle {\delta \Phi_{a}({\bf k},\tau) \over \delta \Phi_{b}({\bf k}',\tau')} \right \rangle  \qquad \tau > \tau' \end{eqnarray}

\noindent By solving the following linearized equations for $R_{ab}(|{\bf k}|,\tau,\tau')$ , $G_{ab}(|{\bf k}|,\tau,\tau')$

\begin{eqnarray} & & \Lambda_{ab}({\bf k}, \tau) R_{bc}(|{\bf k}|,\tau,\tau') =  0   \\ & & \Lambda_{ab}({\bf k},\tau) G_{bc}(|{\bf k}|,\tau,\tau') =  0  \end{eqnarray}

\noindent where 

\begin{equation} \Lambda_{ab} = \delta_{ab}{\partial \over \partial \tau} + \Omega_{ab}({\bf k}, \tau) \end{equation} 

\noindent using initial conditions $R_{ab}(|{\bf k}|,\tau_{i},\tau_{i})=P_{ab}(|{\bf k}|,\tau_{i})$, $G_{ab}(|{\bf k}|,\tau_{i},\tau_{i})=\delta_{ab}$,  one can construct the next to leading order corrections to the power spectrum by solving the equation

\begin{eqnarray} \nonumber  \Gamma_{abcd}({\bf k}, \tau)P_{cd}(|{\bf k}|,\tau) = & & \int^{\tau}_{\tau_{i}}d\tau'' M_{as}(k,\tau,\tau'') R_{bs}(k,\tau,\tau'') + \int^{\tau}_{\tau_{i}} N_{as}(k,\tau,\tau'') G_{bs}(k,\tau,\tau'')  \\ & & + S_{as}(k,\tau) P_{sb}(k,\tau) + (a \leftrightarrow b) \end{eqnarray}

\noindent where

\begin{eqnarray} \nonumber & & M_{as}(k,\tau,\tau'') = 4\int {d^{3}{\bf k}' \over (2\pi)^{3}} \gamma_{apq}({\bf k} - {\bf k}', {\bf k}', \tau)\gamma_{lrs}({\bf k}' - {\bf k}, {\bf k}, \tau'') G_{ql}(k',\tau,\tau'') R_{pr}(|{\bf k} - {\bf k}'|,\tau,\tau''), \\ \nonumber & &  N_{as}(k,\tau,\tau'') = 2\int {d^{3}{\bf k}' \over (2\pi)^{3}} \gamma_{apq}({\bf k} - {\bf k}', {\bf k}', \tau)\gamma_{srl}({\bf k} - {\bf k}', {\bf k}, \tau'') R_{ql}(k',\tau,\tau'') R_{pr}(|{\bf k} - {\bf k}'|,\tau,\tau''), \\ \nonumber & & S_{as}(k,\tau,\tau'') = 3\int {d^{3}{\bf k}' \over (2\pi)^{3}}\sigma_{apqs}({\bf k}', - {\bf k}',{\bf k}, \tau)P_{pq}(k'\tau), \end{eqnarray}

\noindent and 

\begin{equation} \Gamma_{abcd} = \delta_{ac} \delta_{bd} {\partial \over \partial \tau} + \delta_{ac} \Omega_{bd}(k,\tau) + \delta_{bd} \Omega_{ac}(k,\tau) .\end{equation}

\noindent It is the solution to the above set of equations that is used in section \ref{sec:3sub}.

\section{\label{sec:a2}Appendix II: Solar system constraint}

Finally, we discuss the origin of the solar system constraint imposed in Figure~\ref{fig:3}. Whilst many different constraints have been imposed on $f(R)$ models, we feel that the solar system bound considered in (for example) \cite{Hu:2007nk} is both robust and model independent, and it is for these reasons that we incorporate it.

We direct the reader to \cite{Hu:2007nk} for the details of the derivation, and simply state the result,

\begin{equation}\label{eq:app10} |f_{R}(R_{\rm s})| \lesssim 5 \times 10^{-11} \end{equation} 

\noindent where $R_{\rm s} = 8\pi G \rho_{\rm s}$, and $\rho_{\rm s}$ is the typical background density at which solar system tests are performed at. We take this conservatively to be $R_{\rm s} =10^{5} H_{0}^{2}$ \cite{Lin:2010hk,Faulkner:2006ub}, although a thorough treatment would require a detailed description of the solar system density distribution.

It is straightforward to convert this bound into a constraint on the $f(R)$ model parameters for the function ($\ref{eq:i1}$); we find

\begin{equation} |f_{R}(R_{\rm s})| = 2\lambda s \left({R_{\rm vac} \over R_{\rm s}}\right)^{2s+1} \simeq 2\lambda s 10^{-5(2s+1)} < 5 \times 10^{-11}. \end{equation}

\noindent  However for the model ($\ref{eq:107}$) we have not explicitly written a functional form $f(R)$. Therefore to apply the bound ($\ref{eq:app10}$) we must first calculate the relationship between $M(a)$ and $f_{R}$. Using the $a \ll 1$ limit of ($\ref{eq:107}$), rearranging the expression ($\ref{eq:mass}$) and integrating, we find

\begin{equation} \label{eq:ap12} f_{R} = -3 \Omega_{\rm m0} \mu^{2}\int_{0}^{a} \bar{a}^{6\nu-4} d\bar{a} = -{\Omega_{\rm m0} \mu^{2} \over 2\nu-1} a^{6\nu-3} . \end{equation}

\noindent In addition, we require the value of the scale factor at which $R_{\rm cos} = R_{\rm s} \simeq 10^{5} R_{\rm vac}$,

\begin{equation} {R_{\rm s} \over R_{\rm vac} } \simeq 10^{5} =  {\Omega_{m0} \over 4\Omega_{\Lambda}a_{\rm s}^{3}} .\end{equation}

\noindent Hence $a_{s} = (\Omega_{m0} /4\Omega_{\Lambda})^{1/3}10^{-5/3}$. Substituting this into ($\ref{eq:ap12}$) we find that the solar system constraint for the model ($\ref{eq:107}$),

\begin{equation} {\Omega_{\rm m0} \mu^{2} \over 2\nu-1} \left({\Omega_{m0} \over 4\Omega_{\Lambda}}\right)^{2\nu-1}10^{-5(2\nu-1)} < 5 \times 10^{-11} \end{equation}

\end{document}